\documentclass[final,3p,twocolumn,authoryear]{elsarticle}
\usepackage{amssymb}
\journal{Icarus}

\begin{document}
\begin{frontmatter}
\title{Potential Jupiter-Family Comet Contamination of the Main Asteroid Belt}
\author[psi,asiaa]{Henry H.\ Hsieh}\ead{hhsieh@psi.edu}
\author[uh]{Nader Haghighipour}\ead{nader@ifa.hawaii.edu}
\address[psi]{Planetary Science Institute, 1700 E. Ft. Lowell Road, Suite 106, Tucson, AZ 85719, USA}
\address[asiaa]{Institute of Astronomy and Astrophysics, Academia Sinica, P.O. Box 23-141, Taipei 10617, Taiwan}
\address[uh]{Institute for Astronomy, University of Hawaii, 2680 Woodlawn Drive, Honolulu, HI 96822, USA}


\begin{abstract}
We present the results of ``snapshot'' numerical integrations of test particles representing comet-like and asteroid-like objects in the inner solar system aimed at investigating the short-term dynamical evolution of objects close to the dynamical boundary between asteroids and comets as defined by the Tisserand parameter with respect to Jupiter, $T_J$ (i.e., $T_J$$\,=\,$3).  As expected, we find that $T_J$ for individual test particles is not always a reliable indicator of initial orbit types.  Furthermore, we find that a few percent of test particles with comet-like starting elements (i.e., similar to those of Jupiter-family comets) reach main-belt-like orbits (at least temporarily) during our 2~Myr integrations, even without the inclusion of non-gravitational forces, apparently via a combination of gravitational interactions with the terrestrial planets and temporary trapping by mean-motion resonances with Jupiter. We estimate that the fraction of real Jupiter-family comets occasionally reaching main-belt-like orbits on Myr timescales could be on the order of $\sim\,$0.1-1\%, although the fraction that remain on such orbits for appreciable lengths of time is certainly far lower.  For this reason, the number of JFC-like interlopers in the main-belt population at any given time is likely to be small, but still non-zero, a finding with significant implications for efforts to use apparently icy yet dynamically asteroidal main-belt comets as tracers of the primordial distribution of volatile material in the inner solar system.  The test particles with comet-like starting orbital elements that transition onto main-belt-like orbits in our integrations appear to be largely prevented from reaching low eccentricity, low inclination orbits, suggesting that the real-world population of main-belt objects with both low eccentricities and inclinations may be largely free of this potential occasional Jupiter-family comet contamination.  We therefore find that low-eccentricity, low-inclination main-belt comets may provide a more reliable means for tracing the primordial ice content of the main asteroid belt than the main-belt comet population as a whole.  
\end{abstract}


\end{frontmatter}

\section{INTRODUCTION}

\subsection{The Tisserand Parameter}
\label{section:tisserandparam}

The Tisserand parameter, $T_P$, or Tisserand invariant, of a small solar system body under the influence of gravity from the Sun and a major planetary perturber is defined by
\begin{equation}
T_P = {a_P\over a} + 2 \cos i \left[{\left(1-e^2\right){a\over a_P}}\right]^{1/2}
\end{equation}
where $a_P$ is the semimajor axis of the planetary perturber, and $a$, $e$, and $i$ are the semimajor axis, eccentricity, and inclination of the small body in question.  Derived from Jacobi's integral, the long-term value of this quantity is largely conserved in the restricted three-body problem \citep{tis96,vag73}, even in the event of close encounters with the planetary perturber \citep{car95}.

In the study of small solar system body dynamics, the Tisserand parameter with respect to Jupiter, $T_J$, is frequently employed as a discriminant between asteroids and comets.  Main-belt asteroids typically have $T_J$$\,>\,$3, and comets typically have $T_J$$\,<\,$3 \citep{kre72}.
However, despite the appealing simplicity of a clear-cut boundary between asteroids and comets at $T_J$$\,=\,$3, $T_J$ is well-known to be an inexact means of dynamically classifying real solar system objects.
The expression for $T_P$ is derived using an idealized physical approximation in which the orbit of the planetary perturber is assumed to be circular ($e$$\,=\,$0) and non-inclined ($i$$\,=\,$0$^{\circ}$), but Jupiter's actual orbit has both non-zero $e$ and non-zero $i$ ($e_J$$\,=\,$0.0489; $i_J$$\,=\,$1.304$^{\circ}$).
Furthermore, while Jupiter is the dominant planetary perturber in the solar system, the other outer planets as well as the terrestrial planets can also affect cometary orbits \citep[e.g.,][]{mor99,lev06,gal14}.  Lastly, non-gravitational forces such as the Yarkovsky effect \citep[cf.][]{rub95} and cometary outgassing \citep[cf.][]{mar73,yeo04} can potentially play a significant role in the dynamical evolution of solar system objects, such as in the case of 2P/Encke \citep[e.g.,][]{ste96,jfer02,pit04}, but are unaccounted for in the formulation of $T_P$.

\subsection{$T_J$ as an Asteroid-Comet Discriminant}

\citet{fer01,fer05} demonstrated that near-Earth objects (NEOs) with $T_J$$\,<\,$3 (sometimes referred to as asteroids in cometary orbits, or ACOs, in the literature) showed significantly lower albedos than NEOs with $T_J$$\,>\,$3, consistent with the low-$T_J$ objects being dormant or extinct comet nuclei.  This finding led \citet{bin04} and \citet{dem08} to use a combination of low albedos and low $T_J$ values to identify extinct comet candidates in other NEO surveys.  \citet{bin04} reported, however, that dynamical models indicated that $\sim\,$35\% of low-albedo $T_J$$\,\leq\,$3 NEOs were likely to originate from the outer asteroid belt, not the outer solar system as their $T_J$ values might normally suggest.  \citet{zif05} also found that near-infrared spectra of two asteroids with $T_J$$\,<\,$3 showed that they had more in common with X-type asteroids than cometary nuclei.  In a study of asteroids with $T_J$$\,<\,$3, \citet{lic06} similarly found a reflectivity gradient distribution more consistent with outer main-belt asteroids than with cometary nuclei, though cautioned that their results were preliminary.

\begin{figure*}
 \centering{\includegraphics[width=4.0in]{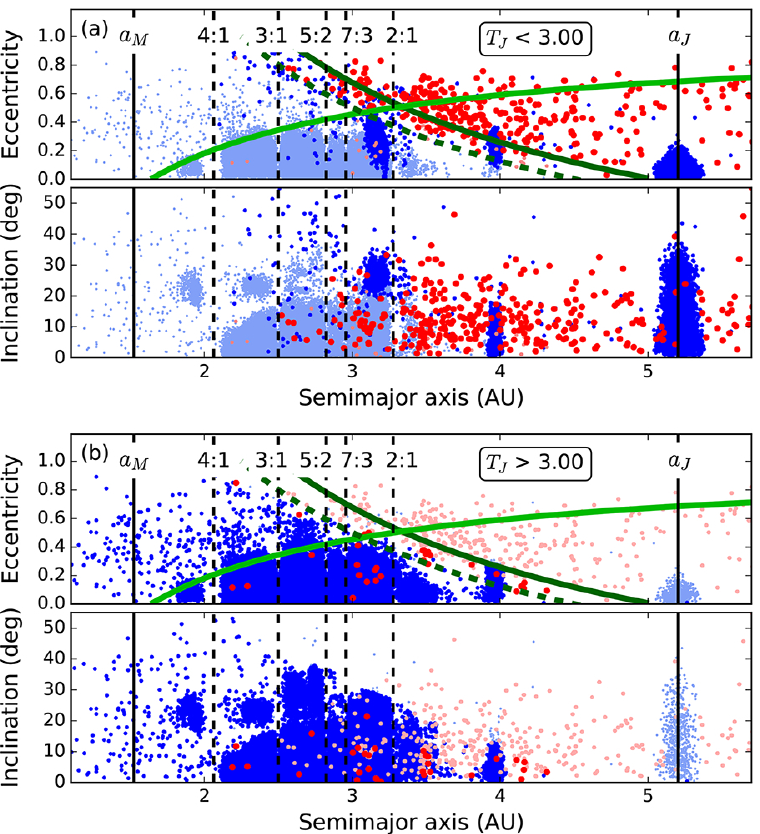}}
\caption{\small Plots of $a$ vs.\ $e$ (top half of each panel) and $i$ (bottom half of each panel) for the first 50\,000 numbered asteroids (pale blue dots) and all comets catalogued by the Minor Planet Center as of 2014 April 1 (pale red dots), where asteroids and comets with $T_J$ values of (a) $T_J$$\,<\,$3.00, and (b) $T_J$$\,>\,$3.00 are highlighted with dark blue and dark red dots, respectively.  Solid vertical lines mark $a$ for Mars and Jupiter ($a_M$ and $a_J$), while the 4:1, 3:1, 5:2, 7:3, and 2:1 MMRs with Jupiter are marked with dashed vertical lines.  The loci of Mars-crossing orbits (where $q$$\,=\,$$Q_{M}$) and Jupiter-crossing orbits (where $Q$$\,=\,$$q_{J}$) are marked with light green and dark green curved solid lines, respectively, on each $a$-$e$ plot, while the loci of orbits for which objects can potentially come within 1.5 Hill radii of Jupiter ($Q$$\,=\,$$q_{J}$$\,-\,$$1.5R_H$) are marked with dark green dashed lines.
}
\label{figure:asteroids_comets_tiss}
\end{figure*}

While physical studies indicate that $T_J$ is probably a reasonable first-order indication of an object's probable dynamical origin, the aforementioned caveats mean that it should not be regarded as an absolute criterion.  Plots of the orbital elements of the first 50\,000 numbered asteroids and all comets catalogued by the Minor Planet Center as of 2014 April 1 (Figure~\ref{figure:asteroids_comets_tiss})
shows that most asteroids have $T_J$$\,>\,$3, and most comets have $T_J$$\,<\,$3, though there are some asteroids with $T_J$$\,<\,$3 and a handful of comets with $T_J$$\,>\,$3.
In particular, Jovian Trojan asteroids ($a$$\,\sim\,$5.2~AU) and Hilda asteroids ($a$$\,\sim\,$3.9$\,-\,$4.0~AU) comprise a large portion of the $T_J$$\,<\,$3 asteroid population.  There are also some $T_J$$\,<\,$3 asteroids within the $a$ bounds of the main asteroid belt (between
the 4:1 and 2:1 mean-motion resonances, or MMRs, with Jupiter), where these objects have larger $e$, larger $i$, or both, relative to the rest of the main-belt asteroid population.  Notably, we see that the vast majority of comets in Figure~\ref{figure:asteroids_comets_tiss} have perihelion distances, $Q$, within 1.5 Hill radii (1.5$R_H$) of Jupiter's perihelion, $q_J$, or beyond, indicating the possibility of very close encounters with the planet and therefore a strong degree of dynamical coupling.  Meanwhile, the vast majority of main-belt asteroids (Hilda and Jovian Trojan asteroids aside) do not have $Q$ closer than 1.5$R_H$ from $q_J$, indicating a low degree of dynamical coupling.  We therefore find that the locus of orbits in $a$-$e$ space with $Q$$\,=\,$$q_J$$\,-\,$$1.5R_H$ forms a reasonably effective alternative dynamical dividing line separating main-belt asteroids and Jupiter-family comets, consistent with the findings of \citet{tan14}.

Continuing to study Figure~\ref{figure:asteroids_comets_tiss}, we see that there are very few comets with $T_J$$\,>\,$3. Most of these have $a$ placing them outside the main asteroid belt (i.e., beyond the 2:1 MMR with Jupiter at 3.277~AU).  A few other comets have $a$ that actually place them within the main asteroid belt, but also have $e$ values larger than those commonly associated with main-belt asteroids.  In almost all of these cases, the orbits of these comets meet the $Q$$\,>\,$$q_J$$\,-\,$$1.5R_H$ criterion for cometary orbits discussed above, with the notable exception of main-belt comets (described below).

\subsection{Main-Belt Comets}

Aside from the aforementioned handful of comets with $T_J$$\,>\,$3 that have $a$ or $e$ placing them beyond the commonly recognized bounds of the main asteroid belt, there exists a newly identified class of comets known as main-belt comets \citep[MBCs;][]{hsi06} that exhibit cometary activity indicative of the sublimation of volatile ices, yet have $T_J$$\,>\,$3, have semimajor axes and eccentricities completely consistent with main-belt asteroids, and do not have close encounters with Jupiter.
MBCs constitute a subset of the group of small solar system bodies known as active asteroids \citep{jew12,jew15}, which also includes disrupted asteroids, which are objects that exhibit comet-like activity that is produced by non-sublimation-driven effects such as impacts or rotational destabilization \citep[cf.][]{hsi12a}.  MBCs are particularly interesting from a dynamical perspective though, since the implication that they are icy bodies raises natural questions about whether they may have originated in the outer solar system like other comets, or whether they were formed in situ as their largely stable main-belt orbits appear to suggest \citep[cf.][]{hsi14}.

Attempts have been made in the past to find plausible dynamical pathways by which Jupiter-family comets (JFCs) could possibly have evolved onto MBC-like orbits, given the unexpectedness of objects on apparently dynamically stable main-belt orbits currently exhibiting active sublimation, but no such pathways were found \citep[e.g.,][]{jfer02}.  The results of numerical integrations attesting to the long-term dynamical stability of individual MBCs \citep{hag09,jew09,hsi12b,hsi12c} appears to indicate that those objects have resided in their current locations in the asteroid belt for some time, and may have even originated there.  There are however a few MBCs which have been found to be unstable on timescales of $\lesssim30$~Myr at their present locations, suggesting that they cannot have resided there for long and must have originated elsewhere \citep[e.g., 238P and 259P;][]{hag09,jew09}.

In this work, we are interested in understanding to what extent $T_J$ and other dynamical characteristics can be used to infer information about an object's possible dynamical origin based on current orbital elements.
The presumed in-situ formation of MBCs is the foundation on which efforts to use them as tracers of primordial ice in the inner solar system are based.  As such, it is very important to determine whether objects currently on main-belt-like orbits can in fact be assumed to be native to the main asteroid belt, or if non-native objects (e.g., from the outer solar system) may be able to occasionally assume main-belt-like orbits, and thus effectively masquerade (at least temporarily) as members of the local native population.  If the latter is the case, identifying the dynamical characteristics of such interlopers would then be very useful for improving our ability to exclude such objects when attempting to infer the distribution and abundance of inner solar system ice from the observed distribution of MBCs.  This issue is of particular interest in astrobiology given that MBCs represent a potential means for constraining solar system formation models that posit that icy objects from the main asteroid belt may be a significant primordial source of Earth's present-day water content.

\section{EXPERIMENTAL DESIGN\label{section:expdesign}}

\begin{figure*}
  \includegraphics[width=6.4in]{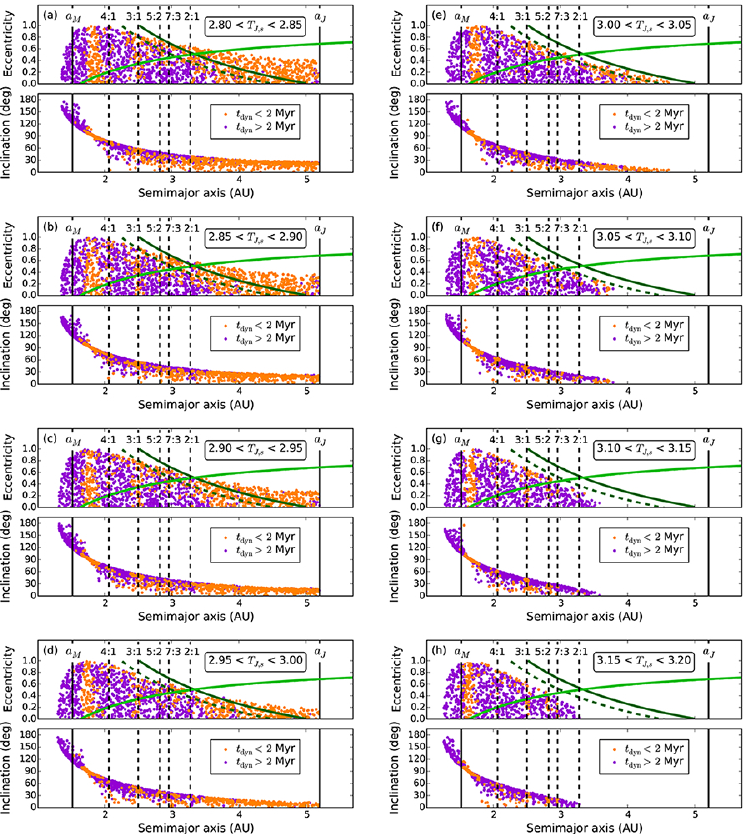}
\caption{\small Plots of $a$ vs.\ $e$ (top half of each panel) and $i$ (bottom half of each panel) for all test particles integrated as part of our study, where test particles with dynamical lifetimes of $t_{\rm dyn}<2$~Myr and $t_{\rm dyn}>2$~Myr are marked with orange and purple dots, respectively, and test particles are separated into individual $T_{J,s}$ bins, as labeled.  Solid vertical lines mark $a_M$ and $a_J$, and the 4:1, 3:1, 5:2, 7:3, and 2:1 MMRs with Jupiter are marked with dashed vertical lines.  The loci of Mars-crossing orbits (where $q=Q_{M}$) and Jupiter-crossing orbits (where $Q=q_{J}$) are marked with light green and dark green curved solid lines, respectively, on each $a$-$e$ plot, while the loci of orbits for which objects can potentially come within 1.5 Hill radii of Jupiter ($Q=q_{J}-1.5R_H$) are marked with dark green dashed lines.
}
\label{figure:testparticles_distribution}
\end{figure*}

For this study, we seek to explore the range of dynamical paths that could be followed by small solar system objects in a designated region of interest in orbital element space (i.e., near the canonical $T_J$$\,=\,$3 boundary between asteroids and comets), with the ultimate objective of determining the degree to which an object's osculating orbital elements (or a parameter derived from those elements, e.g., $T_J$) at some arbitrary point in its dynamical evolution can be relied upon to infer its dynamical history.  To accomplish this, we conducted relatively short-duration integrations of a large number of test particles with starting orbital elements meeting our specified criteria.
Specifically, we generated a sample of 10\,000 test particles spanning a range of $a$, $e$, and $i$ values required to produce starting $T_J$ values ($T_{J,s}$) of 2.80$\,<\,$$T_{J,s}$$\,<\,$3.20, with the expectation that particles close to the canonical dividing line between asteroids and comets should be the most likely to cross that boundary, and thus represent the most interesting cases for study.  To create these test particles, we randomly selected starting $T_{J,s}$, $a_s$, and $e_s$ values from within pre-defined ranges (2.8$\,<\,$$T_{J,s}$$\,<\,$3.2, $a_s$$\,<\,$$a_J$, and 0$\,<\,$$e_s$$\,<\,$0.99, where $a_J$$\,=\,$5.204~AU), and then computed the corresponding $i_s$ value needed to produce the selected $T_{J,s}$ value for each test particle.  Sets of $a_s$ and $e_s$ for which no value of $i_s$ could produce the target $T_{J,s}$ value were discarded and regenerated.  Finally, random values between $0^{\circ}$ to $360^{\circ}$ were selected as arguments of perihelion, longitudes of ascending nodes, and mean anomalies.  The starting orbital elements of all of our test particles generated in this way, separated into individual $T_{J,s}$ bins for added clarity, are plotted in Figure~\ref{figure:testparticles_distribution}.

Of course, by generating test particles that span the entire region of orbital element space where 2.80$\,<\,$$T_{J,s}$$\,<\,$3.20, we sample portions of orbital element space that are only sparsely populated, if at all, by real solar system objects, as can be seen by comparing Figures~\ref{figure:asteroids_comets_tiss} and \ref{figure:testparticles_distribution}.  In particular, our initial test particle set includes objects on polar orbits (i.e., $i_s$$\,\sim\,$90$^{\circ}$) and retrograde orbits (i.e., with $i_s$$\,>\,$90$^{\circ}$).  Our goal in performing this general study, though, is to fully explore the available parameter space to search for possible dynamical pathways for particles defined by a specific dynamical criterion (i.e., $T_{J}$), and investigate what additional dynamical criteria, if any, are able to specify or exclude particular pathways of interest.

We conducted ``snapshot'' integrations of all test particles by integrating each of their orbits forward (using 10-day timesteps) for 2~Myr using the Bulirsch-St\"oer integrator in the Mercury numerical integration software package \citep{cha99}.  The length of our snapshot integrations was chosen so that our study would not require an unmanageably large expenditure of computing resources, while still producing physically meaningful results.
\citet{lev94} found a median dynamical lifetime of $4.5\times10^5$~years for short-period comets before ejection from the solar system or collision with the Sun, and so our integration period of 2~Myr should extend past the dynamical lifetimes of most of the comet-like particles in our integrations.

\begin{figure*}
  \includegraphics[width=6.4in]{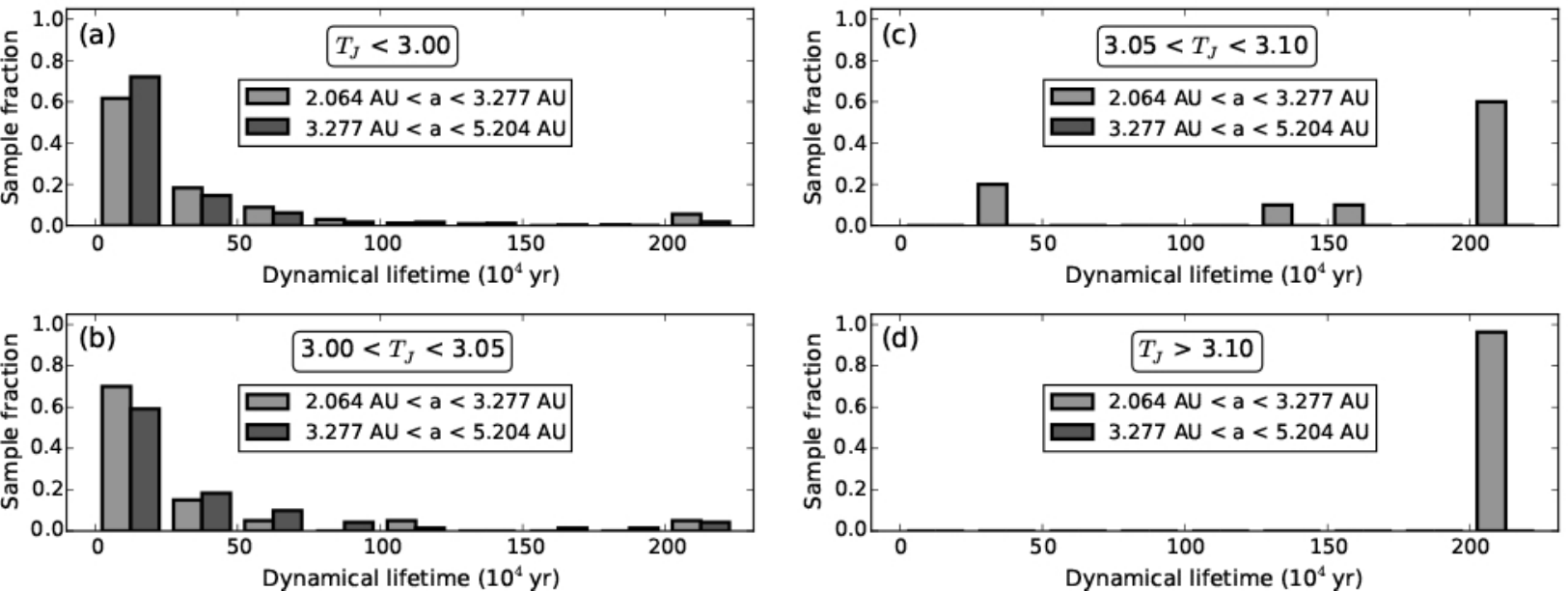}
\caption{\small Histograms of dynamical lifetimes for test particles representing clones of known comets with $a\leq5.204$~AU with $T_{J,s}$ values of (a) $T_{J,s}<3.00$, (b) $3.00<T_{J,s}<3.05$, (c) $3.05<T_{J,s}<3.10$, and (d) $T_{J,s}>3.10$, where light grey bars indicate the fraction of comets with $2.064~{\rm AU} < a < 3.277$~AU that are lost due to ejection or planetary/solar impact within a particular time interval, and dark grey bars indicate the fraction of comets with $a>3.277$~AU that are lost due to ejection or planetary/solar impact within a particular time interval.
}
\label{figure:timeejec_comets}
\end{figure*}

Of course, our chosen integration length means that our results are not appropriate for characterizing the evolution of test particles over much longer timescales, such as the age of the solar system, but this does not mean they are not physically meaningful.  Since we are considering the dynamical evolution of fictitious test particles distributed throughout our orbital element space region of interest in this study, our test particles can be thought of as both representing different individual objects, and also different stages of the long-term dynamical evolution of single objects, hence our characterization of these integrations as snapshot integrations.  Considering the situation from another perspective, longer integrations would certainly more directly characterize the long-term dynamical behavior of our test particles, but in no cases would they be expected to exclude or prevent any dynamical behavior observed during shorter integration periods.  As such, our snapshot integrations can be considered entirely applicable to the study of both the short-term and long-term dynamical evolution of our test particles.


For these integrations, we treated the Sun and the eight major planets as massive bodies (where the mass of Mercury was added to that of the Sun) and all test particles as massless bodies.  Non-gravitational effects were not included.

For reference, we used the same experimental design as we used for our test particles to study the dynamical evolution of known comets.  Since our interest is in comets in the inner solar system, we restrict our sample to those comets with $a<a_{J}$.  We created a set of test particles with the orbital elements of the 639 comets catalogued by the MPC as of 2014 January 1 that met our orbital criteria.  We also created four additional dynamical clones per comet to account for possible chaotic effects due to orbital element uncertainties, giving a total of 3195 cometary test particles and dynamical clones.  For the latter task, we utilized code that generates an arbitrary number of clones of an object that are Gaussian-distributed in orbital element space, centered on the object's osculating orbital elements, and have a distribution characterized by the 1-$\sigma$ uncertainties of the orbital elements of those objects \citep[previously used by][]{hsi12a,hsi12b,hsi12c,hsi13}.

We plot histograms of the dynamical lifetimes of the comets and their clones in each $T_{J,s}$ bin (Figure~\ref{figure:timeejec_comets}), where the dynamical lifetime of an object is the amount of time it spends in the integration before reaching $a=100$~AU and is considered to have been ejected from the solar system, reaching $e>1$, or colliding with a planet or the Sun.  One caveat is that current orbital elements are not available for all comets. Since our integrations were all run from a single starting epoch, it was therefore necessary to integrate objects with non-updated orbital elements up to the present epoch, and in some cases, objects were eliminated from the integrations before even reaching the current epoch.  For the purposes of this analysis, the dynamical lifetimes for these objects were recorded as being zero years.

\begin{figure*}
\centerline{\includegraphics[width=5.0in]{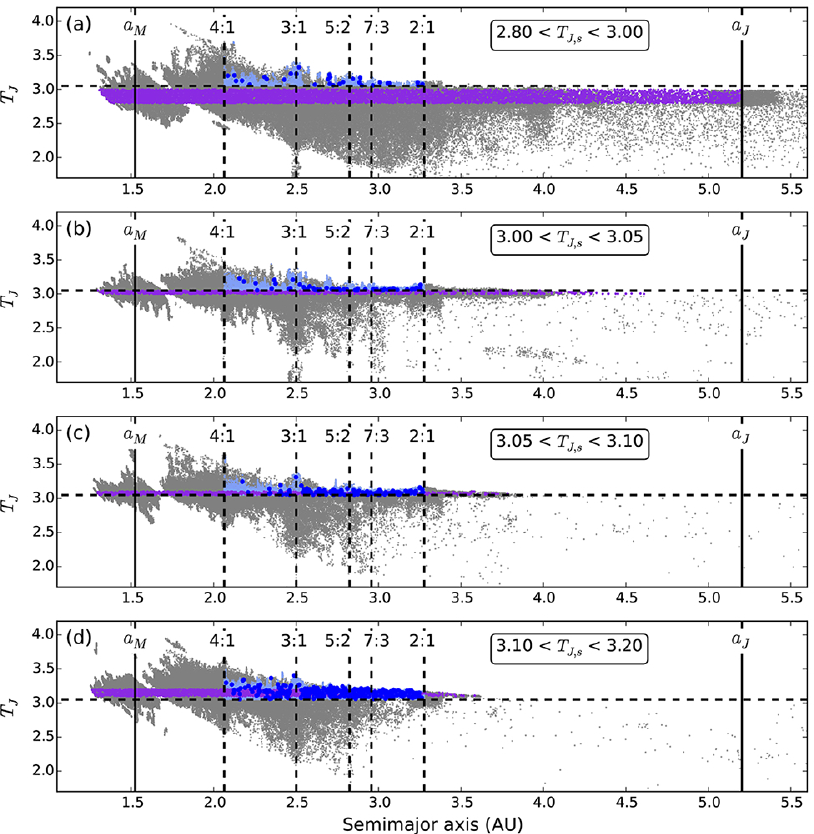}}
\caption{\small Plots of $a$ vs.\ $T_{J,i}$ (small grey dots) for test particles with (a) $2.80<T_{J,s}<3.00$, (b) $3.00<T_{J,s}<3.05$, (c) $3.05<T_{J,s}<3.10$, and (d) $3.10<T_{J,s}<3.20$.  SOEs are plotted with purple dots, while main-belt-like IOEs and FOEs are plotted with light blue and dark blue dots, respectively.  Solid vertical lines mark $a_M$ and $a_J$ in each panel, while the 4:1, 3:1, 5:2, and 2:1 MMRs with Jupiter (from left to right) are marked with dashed vertical lines.  The $T_J=3.05$ boundary is marked by horizontal dashed lines in each panel.
}
\label{figure:semimaj_tiss_tji_all}
\end{figure*}

Overall, we find that $>$95\% of our comet test particles are lost prior to the end of our 2~Myr integrations (consistent with the results of integrations of JFCs by \citet{jfer02} over an identical integration period), indicating that our integrations are indeed longer than the dynamical lifetimes of most comets in the inner solar system.  We also immediately see that the dynamical behavior of comets with $3.00<T_{J,s}<3.05$ is extremely similar to those with $T_{J,s}<3.00$, indicating that $T_J=3$ may not in fact be an appropriate dividing line between comet-like and asteroid-like orbits. A shift towards greater stability is seen for comet test particles with $3.05<T_{J,s}<3.10$, while all comets with $T_{J,s}>3.10$ are found to be stable over the entirety of our integrations.  No appreciable differences in dynamical lifetimes were seen for comets with semimajor axes interior to and exterior to the 2:1 MMR with Jupiter (i.e., the outer boundary of the main asteroid belt).
The short ($<1$~Myr) dynamical lifetime of a typical comet is a key dynamical characteristic indicating a likely recent insertion onto an inner-solar-system-crossing orbit given the low likelihood of it residing on that orbit for significantly longer than its calculated dynamical lifetime.  As such, hereafter, we will consider $T_J=3.05$ to be, in practice, a more appropriate approximate upper bound on ``comet-like'' orbits.  This is consistent with the modified $T_J$ criterion for differentiating asteroids and comets used by \citet{tan14}, as well as our previous discussion of how the physical simplifications used to derive $T_J$ are inexact (Section~\ref{section:tisserandparam}).

\section{RESULTS \& ANALYSIS\label{results}}

\subsection{Reliability of $T_J$ as a Dynamical Discriminant\label{section:tisserand_reliability}}

\begin{figure}
\centerline{\includegraphics[width=3in]{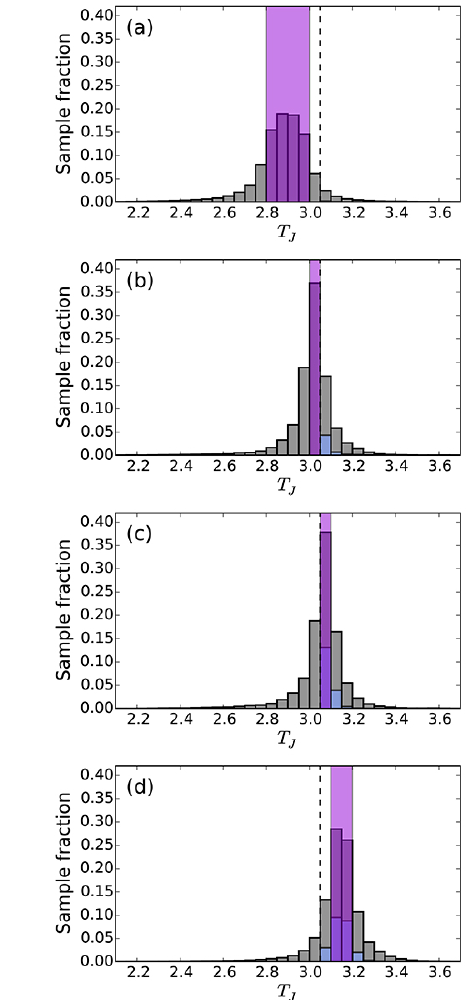}}
\caption{\small Histograms of $T_{J,i}$ values (grey bars) for test particles with (a) $2.80<T_{J,s}<3.00$, (b) $3.00<T_{J,s}<3.05$, (c) $3.05<T_{J,s}<3.10$, and (d) $3.10<T_{J,s}<3.20$.  The ranges of $T_{J,s}$ values are marked with shaded purple regions, while the $T_J$ distribution of main-belt-like IOEs are over-plotted as light blue bars.  The $T_J=3.05$ boundary is marked by vertical dashed lines in each panel.
}
\label{figure:histograms_tji_all}
\end{figure}

In order to investigate the degree to which $T_J$ remains a reliable dynamical parameter for identifying the origins of individual test particles, even over relatively short periods of time, we compare the starting orbital elements (SOEs), intermediate orbital elements (IOEs) in 10\,000-year intervals over our entire 2~Myr integration period or until particles are lost due to ejection or a planetary or solar impact, and final orbital elements (FOEs) of our test particles.  We plot starting, intermediate, and final $a$ and $T_J$ values for test particles in different $T_{J,s}$ bins in Figure~\ref{figure:semimaj_tiss_tji_all}.  We immediately see from these plots that even $T_J$$\,=\,$3.05 is not a particularly impenetrable dynamical boundary, with a significant fraction of particles with starting $T_J$ values, $T_{J,s}$, of 2.80$\,<\,$$T_{J,s}$$\,<\,$3.00 reaching intermediate $T_J$ values ($T_{J,i}$) of $T_{J,i}$$\,>\,$3.05 over the course of the integration, and particles with 3.10$\,<\,$$T_{J,s}$$\,<\,$3.20 also reaching $T_{J,i}$$\,<\,$3.05.  Notably, some particles with 2.80$\,<\,$$T_{J,s}$$\,<\,$3.00 even attain main-belt-like IOEs\footnote{For the purposes of the analyses presented here, we define ``main-belt-like'' orbits as those having $T_J$$\,>\,$3.05 (i.e., dynamically decoupled from Jupiter), $2.064~{\rm AU}<a<3.277~{\rm AU}$ (i.e., $a$ between the 4:1 and 2:1 MMRs with Jupiter, which bound the canonical main asteroid belt region), and $q>(Q_{M}+1.5R_{H,{M}})$ and $Q<(q_{J}-1.5R_{H,{J}})$ (i.e., confined within the orbits of Mars and Jupiter and prevented from approaching within 1.5$R_H$ of either planet), where $(Q_{M}+1.5R_{H,{M}})=1.65$~AU and $(q_{J}-1.5R_{H,{J}})=4.50$~AU.  ``Comet-like'' orbits are defined as those having $T_J$$\,<\,$3.05 and $Q>(q_{J}-1.5R_{H,{J}})$ (i.e., dynamically coupled to Jupiter).  In all cases, descriptions of orbits as ``main-belt-like'' or ``comet-like'' are intended only to refer to an object's or test particle's orbital elements at a particular moment in time, and are not meant to imply anything further about the object's dynamical history, long-term stability, or evolutionary fate.} during the integration period, and some even have main-belt-like FOEs at the end of the integrations.

\begin{table*}[ht]
\normalsize
\caption{\normalsize Distribution of $T_{J,i}$ Values}
\smallskip
\begin{tabular}{ccccc}
\hline\hline
 \multicolumn{1}{c}{$T_{J,s}$ Bin} &
 \multicolumn{1}{c}{$T_{J,i}<3.00^a$} &
 \multicolumn{1}{c}{$T_{J,i}<3.05^b$} &
 \multicolumn{1}{c}{$T_{J,i}>3.05^c$} &
 \multicolumn{1}{c}{$T_{J,i}>3.10^d$} \\
\hline
$T_{J,s}<3.00$      & 0.87 & 0.93 & 0.07 & 0.04 \\
$3.00<T_{J,s}<3.05$ & 0.34 & 0.71 & 0.29 & 0.12 \\
$3.05<T_{J,s}<3.10$ & 0.16 & 0.35 & 0.65 & 0.27 \\
$T_{J,s}>3.10$      & 0.07 & 0.12 & 0.88 & 0.74 \\
\hline\hline
\end{tabular}
\newline{$^a$ Fraction of IOEs where $T_{J,i}<3.00$.}
\newline{$^b$ Fraction of IOEs where $T_{J,i}<3.05$.}
\newline{$^c$ Fraction of IOEs where $T_{J,i}>3.05$.}
\newline{$^d$ Fraction of IOEs where $T_{J,i}>3.10$.}
\label{table:intermediate_tj_dist}
\end{table*}

In Figure~\ref{figure:histograms_tji_all}, we plot histograms of all $T_{J,i}$ values attained by test particles in each $T_{J,s}$ bin in order to further investigate their distribution.  In Table~\ref{table:intermediate_tj_dist}, we also list the fractions of $T_{J,i}$ values on either side of the ostensible $T_J=3.05$ asteroid-comet boundary attained by test particles in each $T_{J,s}$ bin.  While the majority of test particles with $T_{J,s}<3.00$ remain below the $T_J=3.05$ boundary throughout the integration period, some do reach $T_{J,i}>3.05$ (and even $T_{J,i}>3.10$) for at least a portion of the time covered by our integrations.  Similarly, while the majority of test particles with $T_{J,s}>3.10$ remain above the $T_J=3.05$ boundary, some reach $T_{J,i}<3.05$ (and even $T_{J,i}<3.00$) during a portion of the integration period.  By comparison, test particles with $3.00<T_{J,s}<3.10$ spend substantial amounts of time on the opposite side of the $T_J=3.05$ boundary from where they originated.

\subsection{Transfer of Objects from Comet-Like Orbits to Main-Belt Orbits\label{section:mbc_origins}}

\subsubsection{Results of Initial Integrations\label{section:mbc_origins_initial}}

\begin{figure*}
\centerline{\includegraphics[width=6.5in]{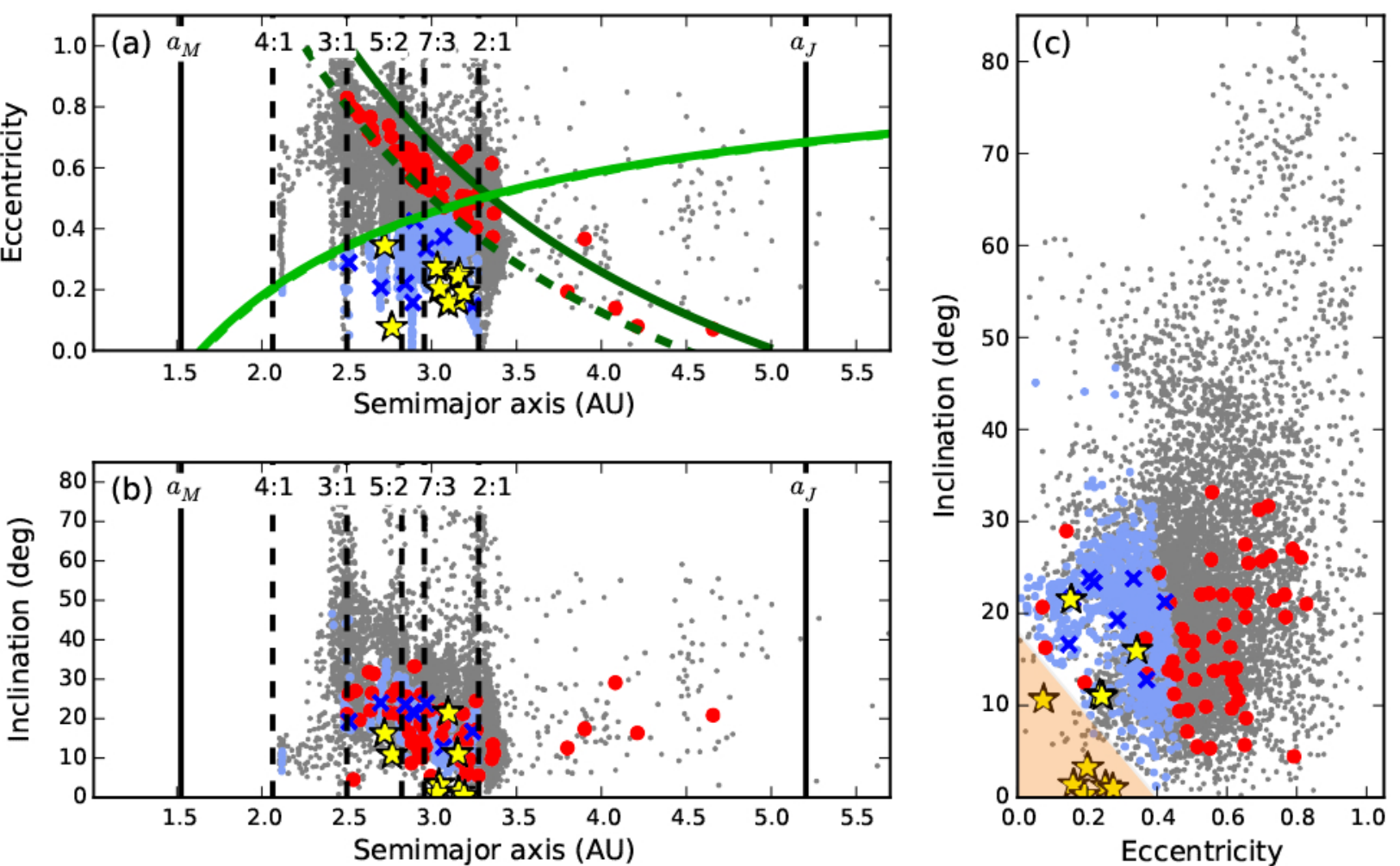}}
\caption{\small Plots of (a) $a$ vs.\ $e$, (b) $a$ vs.\ $i$, and (c) $e$ vs.\ $i$ in 10\,000-year intervals for test particles that have comet-like SOEs and that reach main-belt-like IOEs at any point during the integration period.  SOEs are plotted with red dots, while main-belt-like IOEs and FOEs that meet main-belt criteria are plotted with light blue dots and dark blue X's, respectively.  All other IOEs are plotted with small grey dots.  Orbital elements of the known MBCs are plotted with yellow stars.  In (a) and (b), $a_M$ and $a_J$ are marked with solid vertical lines, while the semimajor axes of the 4:1, 3:1, 5:2, 7:3, and 2:1 MMRs with Jupiter are marked with dashed vertical lines. In (a), the loci of Mars-crossing orbits (where $q=Q_{M}$) and Jupiter-crossing orbits (where $Q=q_{J}$) are marked with light green and dark green curved solid lines, respectively, and the loci of orbits for which objects can potentially come within 1.5$R_H$ of Jupiter ($Q=q_{J}-1.5R_H$) are marked with a dark green dashed line.  In (c), the approximate region of $e$-$i$ space into which comet-like test particles never enter is shaded in orange.
}
\label{figure:start_cometlike_sometimes_mb_all}
\end{figure*}

In Figure~\ref{figure:start_cometlike_sometimes_mb_all}, we plot IOEs of test particles with comet-like SOEs that attain main-belt-like IOEs at any time during the integration period, as well as the current orbital elements of the known MBCs.  As seen before, we find that test particles with comet-like SOEs reach a significant portion of main-belt orbital element space.   Specifically, of the 1727 test particles in our integrations that had comet-like SOEs, 57 test particles ($\sim$3.5\% of the total sample) reached main-belt-like orbits at some point during the integrations, while 8 of those test particles actually had main-belt-like FOEs.  We caution that since our sample of comet-like test particles is not a realistic representation of the known comet population, these rates are not expected to accurately reflect the real-world rates of JFCs evolving onto main-belt-like orbits.

We note that 29 clones of real comets with $T_{J,i}$$\,<\,$3.05 (out of 2630 such clones, or $\sim\,$1\%) also attained main-belt-like IOEs at some point when integrated for 2~Myr (Section~\ref{section:expdesign}), and two of those objects (clones of 249P and P/2005 JQ5; $\sim$0.1\% of the total sample of comet clones) had main-belt-like FOEs.  Like our test particle set, due to historical discovery biases, this set of comet clones likewise does not necessarily accurately represent the current steady-state population of JFCs.  Nonetheless, we conclude from these results that the fraction of comet-like objects that may attain main-belt-like orbits at some point during their dynamical lifetimes, albeit perhaps only temporarily, is non-zero, and estimate that it may be on the order of $\sim$0.1-1\%.

That said, we see that in our test particle integrations, particles with comet-like SOEs that reach main-belt-like orbits do not attain very low $e$ and very low $i$ simultaneously.  We can approximately define a region of $e$-$i$ space, shaded in orange in Figure~\ref{figure:start_cometlike_sometimes_mb_all}, into which test particles with comet-like SOEs do not enter over the course of our integrations, and therefore appears to be ``protected'' from comet-like interlopers.  This region is approximately empirically defined by
\begin{equation}
0.775e+\sin(i)<0.31
\label{equation:protected_region}
\end{equation}
and includes the current orbital elements of (1) Ceres, 133P, 176P, 238P, 288P, and P/2013 R3.

\begin{figure*}
\centerline{\includegraphics[width=4.3in]{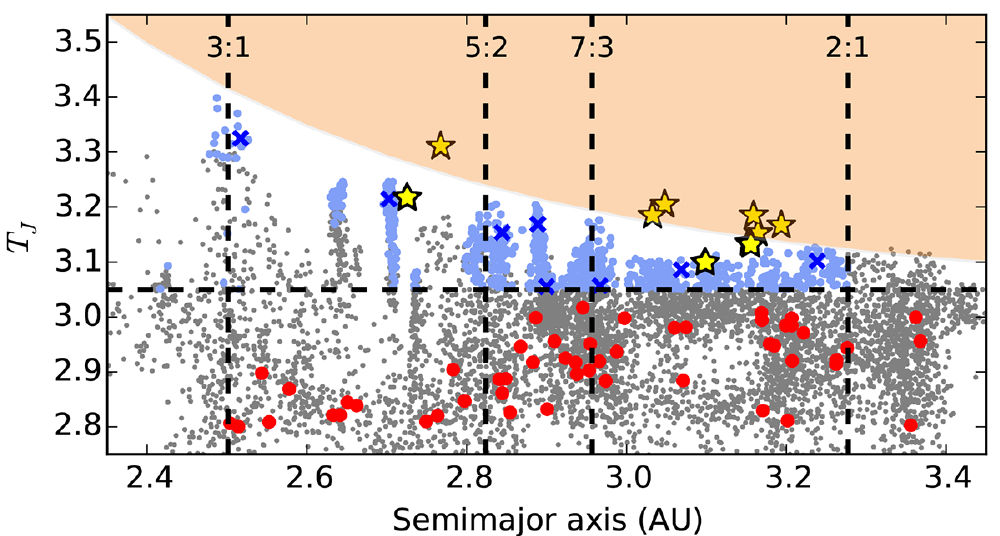}}
\caption{\small Plot of $a$ vs.\ $T_J$ in 10\,000-year intervals for test particles that have comet-like SOEs and that reach main-belt-like IOEs at any point during the integration period.  SOEs are plotted with red dots, while main-belt-like IOEs and FOEs that meet main-belt criteria are plotted with light blue dots and dark blue X's, respectively.  All other IOEs are plotted with small grey dots.  Orbital elements of the known MBCs are plotted with yellow stars.  The semimajor axes of the 3:1, 5:2, 7:3, and 2:1 MMRs with Jupiter (from left to right) are marked with dashed vertical lines.  The approximate region of $a$-$T_J$ space into which comet-like test particles never enter, corresponding to the analogous highlighted region of $e$-$i$ space in Figure~\ref{figure:start_cometlike_sometimes_mb_all}, is shaded in orange.
}
\label{figure:start_cometlike_sometimes_mb_atj}
\end{figure*}

For reference, we also plot $a$ vs.\ $T_J$ for the IOEs of the test particles from Figure~\ref{figure:start_cometlike_sometimes_mb_all}, and identify an analogous ``protected'' region in $a$-$T_J$ space into which test particles with comet-like SOEs do not enter (Figure~\ref{figure:start_cometlike_sometimes_mb_atj}).  This region can be approximately empirically described by
\begin{equation}
T_J > 55\cdot\exp({-2a})+3.05
\label{equation:protected_region_atj}
\end{equation}
and contains the orbital elements of the same MBCs in the protected region in $e$-$i$ space identified in Figure~\ref{figure:start_cometlike_sometimes_mb_all}.
Notably, the protected region in $a$-$T_J$ space appears to extend to lower $T_J$ values in the outer main belt than the inner main belt.  Practically speaking, for example, this means that having $T_J$$\,>\,$3.15 is sufficient for an object to be located in the protected region in the outer main belt ($a>3$~AU), while for $a<2.8$~AU, values of $T_J>3.25$ or even higher are needed for an object to be in the protected zone.  This may perhaps be related to weaker dynamical coupling of objects with Jupiter with increasing average distance from the planet, but more detailed theoretical analysis of this issue (that is beyond the scope of the general study presented here) will be needed to ascertain the exact causes of this behavior.

\setlength{\tabcolsep}{5pt}
\begin{table*}[ht]
\normalsize
\caption{\normalsize Test Particles with Comet-Like SOEs and Main-Belt-Like FOEs}
\smallskip
\begin{tabular}{ccccccccccccccc}
\hline\hline
 \multicolumn{1}{c}{Particle} &&
 \multicolumn{1}{c}{$a_s^a$} &
 \multicolumn{1}{c}{$e_s$} &
 \multicolumn{1}{c}{$i_s^b$} &
 \multicolumn{1}{c}{$T_{J,s}$} & &
 \multicolumn{1}{c}{$a_f^c$} &
 \multicolumn{1}{c}{$e_f$} &
 \multicolumn{1}{c}{$i_f^d$} &
 \multicolumn{1}{c}{$T_{J,f}$} &&
 \multicolumn{1}{c}{MMR$^e$} &
 \multicolumn{1}{c}{$a_{\rm MMR}^f$} &
 \multicolumn{1}{c}{$\Delta a_f^g$} \\
\hline
A && 2.505 & 0.826 & 21.018 & 2.806 & & 2.517 & 0.289 & 19.278 & 3.324 & &  3:1 & 2.501 & 0.016 \\ 
B && 2.855 & 0.661 & 25.457 & 2.827 & & 2.703 & 0.208 & 23.910 & 3.214 & &  8:3 & 2.705 & 0.002 \\ 
C && 2.955 & 0.561 & 17.465 & 2.951 & & 2.900 & 0.424 & 21.228 & 3.055 & & 12:5 & 2.902 & 0.002 \\ 
D && 2.911 & 0.593 & 14.106 & 2.956 & & 2.967 & 0.334 & 23.785 & 3.056 & &  7:3 & 2.957 & 0.010 \\ 
E && 2.634 & 0.716 & 31.668 & 2.821 & & 2.844 & 0.221 & 23.369 & 3.153 & &  5:2 & 2.824 & 0.020 \\ 
F && 2.988 & 0.525 & 22.045 & 2.937 & & 2.889 & 0.159 & 21.654 & 3.168 & & 12:5 & 2.902 & 0.013 \\ 
G && 3.075 & 0.470 & 18.323 & 2.981 & & 3.068 & 0.372 & 12.779 & 3.086 & &  9:4 & 3.029 & 0.039 \\ 
H && 3.207 & 0.432 & 13.874 & 2.997 & & 3.239 & 0.149 & 16.636 & 3.101 & &  2:1 & 3.277 & 0.038 \\ 
\hline\hline
\end{tabular}
\newline{$^a$ Starting semimajor axis, in AU.}
\newline{$^b$ Starting inclination, in degrees.}
\newline{$^c$ Final semimajor axis, in AU.}
\newline{$^d$ Final inclination, in degrees.}
\newline{$^e$ Nearest major or moderate-order MMR}
\newline{$^f$ Semimajor axis, in AU, of nearest major or moderate-order MMR}
\newline{$^g$ Absolute value of the difference, in AU, between $a_f$ for each particle and the semimajor axis of the nearest major or moderate-order MMR.}
\label{table:comet_to_mainbelt}
\end{table*}

Specifically examining the eight test particles that are seen to have comet-like SOEs and main-belt-like FOEs, we assign labels (``A''-``H'') to each test particle and list their SOEs and FOEs in Table~\ref{table:comet_to_mainbelt}.  First, we note that in all cases, $a_s$ and $a_f$ for each particle differ by relatively small amounts (five of the eight particles undergo net changes in $a$ of $<0.1$~AU, while three undergo net changes in $a$ of $\sim\,$0.1$\,-\,$0.2~AU). This appears to place a practical limit of $a_s$$\,>\,$2.25~AU for most particles with comet-like SOEs entering the main-belt, considering the initial requirement of $Q_{i}>(q_{J}-1.5R_{H,{J}})$ for a particle to be considered to have comet-like SOEs.
It should also be noted that by only considering test particles with comet-like SOEs and main-belt-like FOEs here, we are focusing on a very select group of test particles that follow a very specific dynamical evolutionary path.  Other particles that experience much larger semimajor axis changes are more likely to be ejected prior to the end of the integrations, leaving just those particles that happen to dynamically evolve in ways that do not change their semimajor axes too drastically.
Meanwhile, substantial decreases in $e$ are seen for all particles highlighted here, while $i$ is seen to vary inconsistently.  For two of the eight particles (E and G), $i$ declines significantly ($>5^{\circ}$) between the beginning and end of our integrations, but for three other particles (A, B, and F), $i$ decreases by less than 2$^{\circ}$, and for the final three particles (C, D, and H), $i$ actually increases.

While each of these particles have $T_{J,s}$$\,<\,$3.00, five of the eight particles have final $T_J$ values ($T_{J,f}$) of $T_{J,f}$$\,>\,$3.10 at the end of our 2~Myr integrations, meaning that $T_{J,f}$$\,\gg\,$$T_{J,s}$ for all of these particles.  This means that all of these particles begin on unambiguously comet-like orbits, and five of the eight particles have transitioned (at least temporarily) to unambiguously main-belt-like orbits at the end of 2~Myr, with only three particles (C, D, and G) ending in the somewhat ambiguous 3.05$\,<\,$$T_J$$\,<\,$3.10 bin between the two extremes (cf.\ Figure~\ref{figure:timeejec_comets}).  Finally, we note that essentially all of these particles have $a_f$ values placing them extremely close to a major, or at least moderate-order (i.e., low-integer), MMR (Table~\ref{table:comet_to_mainbelt}), suggesting that these resonances may be responsible for helping to temporarily trap these objects in the main belt during the integration period, presumably by providing protection against close encounters with Jupiter \citep[e.g.,][]{gla97,mal99,gab03,pit04,bro05,car08,jfer14}.


Intriguingly, some of these eight particles have SOEs similar to those of currently known JFCs, while their FOEs are similar to those of currently known MBCs, suggesting that it may in fact be possible for JFCs to at least occasionally take on MBC-like orbits.  In Table~\ref{table:jfcs_mbcs_highlighted}, we list known JFCs with current orbital elements similar to the SOEs of these test particles, and MBCs with current orbital elements similar to the FOEs of these test particles. These results suggest that there is a non-zero probability that the listed MBCs could have JFC-like origins.  Of the JFCs listed in this table, we note that 197P and one of its dynamical clones do in fact temporarily reach a main-belt-like orbit during our initial 2~Myr test integrations (Section~\ref{section:expdesign}), where the clone remains on such an orbit for almost 1~Myr. However, neither of them remain on a main-belt-like orbit until the end of those integrations.  Of the 2168 comets or dynamical clones of comets with JFC-like SOEs integrated in Section~\ref{section:expdesign}, 15 objects ($<\,$1\% of the total sample) have main-belt-like IOEs at some point during the 2-Myr test integrations, albeit most only very briefly (i.e., for only a few timesteps at a time).  In addition to 197P and one of its clones, other exceptions include one clone each of 249P, P/2004 T1, and P/2005 JQ5, which attain main-belt-like IOEs for relatively long periods of time (i.e., 0.5$\,-\,$1~million yrs), where the clones of 249P and P/2005 JQ5 actually have main-belt-like FOEs.  We emphasize though that having main-belt-like FOEs after just 2~Myr is no guarantee of long-term stability, and further note that none of the actual comets in these cases reach main-belt-like IOEs at any time during the same integrations.

\begin{table*}[ht]
\caption{\small JFCs and MBCs with Orbital Similarities to Particles with Comet-Like SOEs and Main-Belt-Like FOEs}
\smallskip
\small
\begin{tabular}{lcccccccc}
\hline\hline
 \multicolumn{1}{c}{Object} &&
 \multicolumn{1}{c}{$a^a$} &
 \multicolumn{1}{c}{$e$} &
 \multicolumn{1}{c}{$i^b$} &
 \multicolumn{1}{c}{$T_{J}$} &
 \multicolumn{1}{c}{MMR$^c$} &
 \multicolumn{1}{c}{$a_{\rm MMR}^d$} &
 \multicolumn{1}{c}{$\Delta a^e$} \\
\hline
{\it JFCs} \\
197P/LINEAR           && 2.866 & 0.630 & 25.54 & 2.856 & 5:2 & 2.824 & 0.042 \\
189P/NEAT             && 2.921 & 0.597 & 20.38 & 2.909 & 7:3 & 2.957 & 0.036 \\
182P/LONEOS           && 2.931 & 0.666 & 16.91 & 2.846 & 7:3 & 2.957 & 0.026 \\
26P/Grigg-Skjellerup  && 3.017 & 0.640 & 22.43 & 2.806 & 9:4 & 3.029 & 0.012 \\
294P/LINEAR           && 3.200 & 0.595 & 19.09 & 2.818 & 2:1 & 3.277 & 0.077 \\
\hline
{\it MBCs} \\
259P/Garradd          && 2.726 & 0.342 & 15.90 & 3.217 &  8:3 & 2.705 & 0.021 \\
324P/La Sagra         && 3.099 & 0.154 & 21.40 & 3.099 & 13:6 & 3.106 & 0.007 \\
P/2012 T1             && 3.154 & 0.236 & 11.06 & 3.135 &  2:1 & 3.277 & 0.123 \\
313P/Gibbs            && 3.156 & 0.242 & 10.97 & 3.132 &  2:1 & 3.277 & 0.121 \\
\hline\hline
\end{tabular}
\newline{$^a$ Semimajor axis, in AU.}
\newline{$^b$ Inclination, in degrees.}
\newline{$^c$ Nearest major or moderate-order MMR}
\newline{$^d$ Semimajor axis, in AU, of nearest major or moderate-order MMR}
\newline{$^e$ Difference, in AU, between $a_f$ for each object and the semimajor axis of the nearest major or moderate-order MMR.}
\label{table:jfcs_mbcs_highlighted}
\end{table*}

\subsubsection{Detailed Orbital Evolution Analysis\label{section:mbc_origins_detailed}}

\begin{figure*}
\centerline{\includegraphics[width=6.5in]{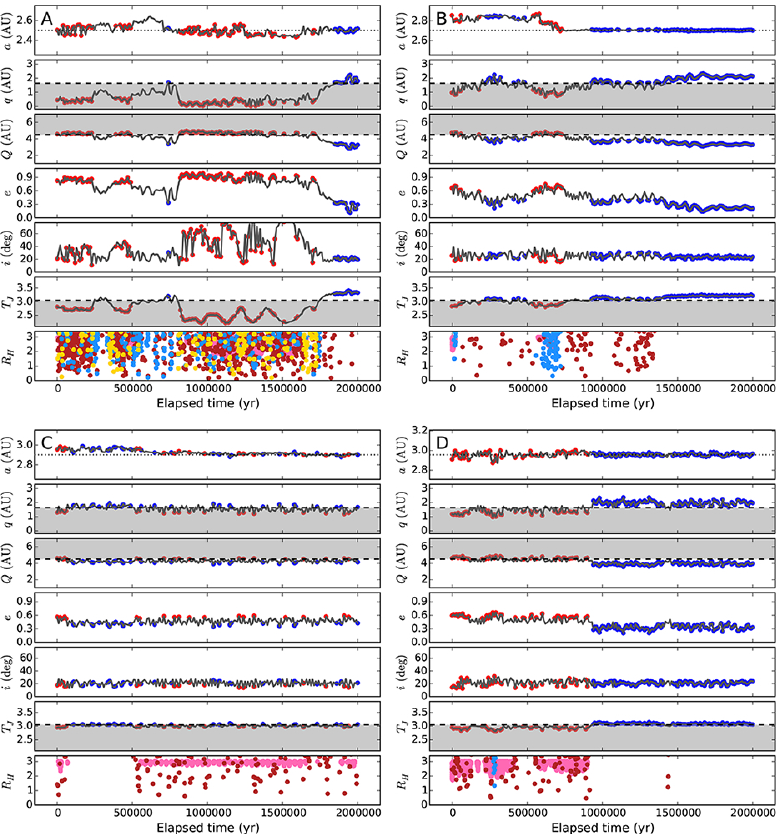}}
\caption{\small Plots of orbital parameter evolution (black lines) over the course of our integrations for particles A-D in Table~\ref{table:comet_to_mainbelt}, as labeled.  Orbital parameters plotted in the first six sub-panels for each particle are, from top to bottom, $a$ (in AU), $q$ (in AU), $Q$ (in AU), $e$, $i$ (in degrees), and $T_J$, where red dots indicate where a particle's orbital elements are comet-like and blue dots indicate where a particle's orbital elements are main-belt-like.  The semimajor axis corresponding to the nearest major or moderate-order MMR to each particle's FOEs (cf.\ Table~\ref{table:comet_to_mainbelt}) is marked by a horizontal dotted line in the first sub-panel for each particle.  The final plot in each panel shows the distances of close planetary encounters in units of $R_H$ for each respective planet, where pink, dark red, light blue, and yellow dots show encounters with Jupiter, Mars, Earth, and Venus, respectively.  Grey shaded regions indicate $a$, $q$, $Q$, and $T_J$ ranges that do not meet main-belt criteria (cf.\ Section~\ref{section:tisserand_reliability}).
}
\label{figure:mbparticles_evolution_ABCD}
\end{figure*}

\begin{figure*}
\centerline{\includegraphics[width=6.5in]{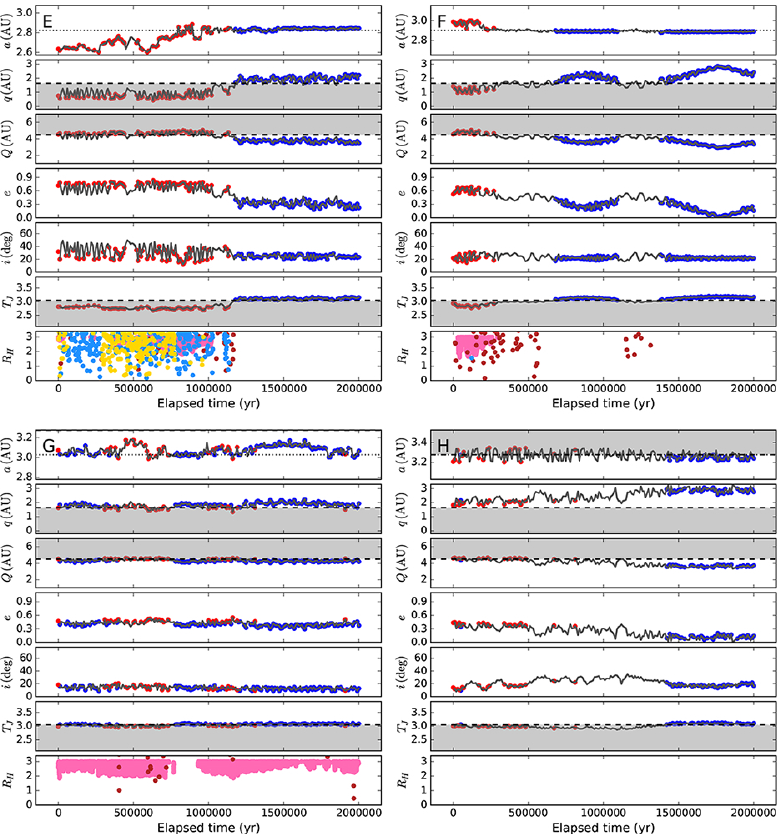}}
\caption{\small Same as Figure~\ref{figure:mbparticles_evolution_ABCD}, but for particles E-H in Table~\ref{table:comet_to_mainbelt}, as labeled.
}
\label{figure:mbparticles_evolution_EFGH}
\end{figure*}

\begin{figure*}
\centerline{\includegraphics[width=6.5in]{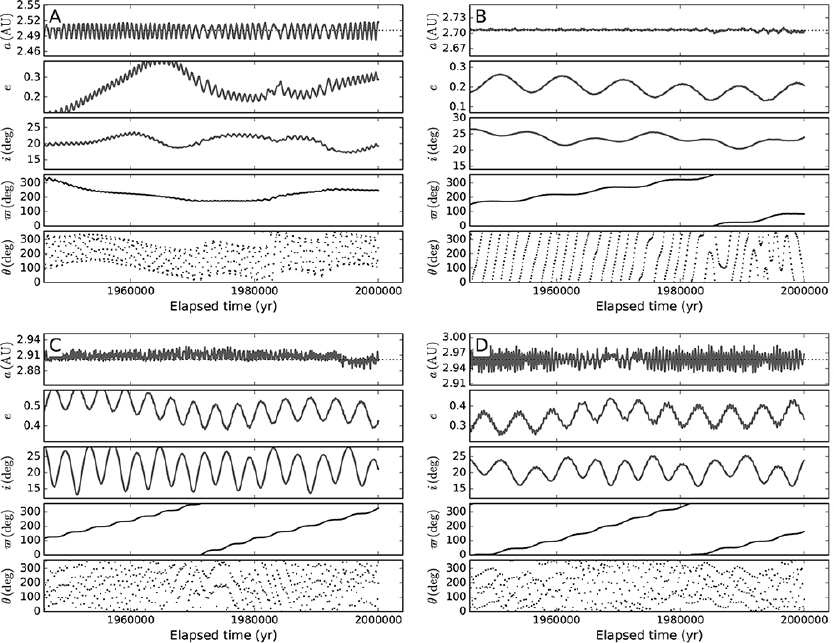}}
\caption{\small Plots of the time evolution of $a$, $e$, $i$, $\varpi$, and the resonant angle, $\theta$, corresponding to the suspected associated MMR for each particle listed in Table~\ref{table:comet_to_mainbelt} in 100~yr intervals over the final $\sim$50\,000 years of our integrations for particles A-D, as labeled.  The semimajor axis corresponding to the nearest major or moderate-order MMR to each particle's FOEs (cf.\ Table~\ref{table:comet_to_mainbelt}) is marked by a horizontal dotted line in the first sub-panel for each particle.
}
\label{figure:mbparticles_evolution_ABCD_100yr}
\end{figure*}

\begin{figure*}
\centerline{\includegraphics[width=6.5in]{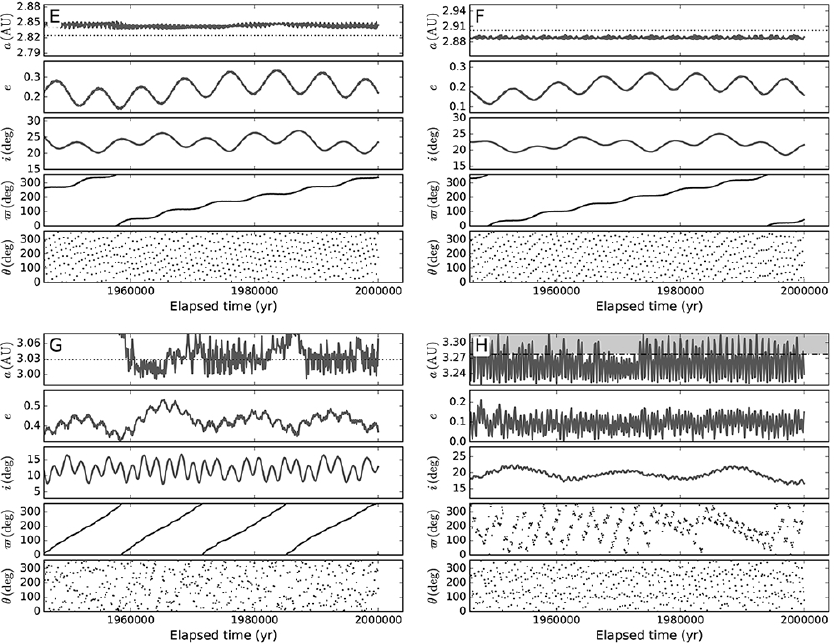}}
\caption{\small Same as Figure~\ref{figure:mbparticles_evolution_ABCD_100yr}, but for particles E-H in Table~\ref{table:comet_to_mainbelt}, as labeled.  Grey shaded regions indicate $a$ ranges that do not meet main-belt criteria (cf.\ Section~\ref{section:tisserand_reliability}).
}
\label{figure:mbparticles_evolution_EFGH_100yr}
\end{figure*}

We plot the evolution of key orbital parameters, as well as the times and distances to planets of encounters at distances of $<\,$$3R_H$ (for each planet's respective $R_H$) for particles A-H over the course of our 2-Myr integrations in Figures~\ref{figure:mbparticles_evolution_ABCD} and \ref{figure:mbparticles_evolution_EFGH}, marking intermediate times at which their orbits are comet-like and main-belt-like.  As we noted in Section~\ref{section:mbc_origins_initial}, $a$ varies by relatively small amounts for each particle over the course of the integrations.  Values of $e$ (and therefore $q$ and $Q$) and $i$, however, are seen to change dramatically for most particles, though the timescales of these changes varies from particle to particle.  Notably, we see that while the semimajor axes of most of these highlighted particles appear to be strongly associated with the nearby MMRs listed in Table~\ref{table:comet_to_mainbelt} over at least some portion of the integrations, some exhibit irregular fluctuations (e.g., particles C, D, and G) or small consistent offsets from the suspected associated MMR (e.g., particles E, F, and H), suggesting that some of these particles are additionally influenced by other nearby and possibly overlapping two- and three-body MMRs \citep[where overlapping MMRs can actually impart additional short-term stability in certain cases;][]{gab03}, or other secular effects.  For reference, we show more detailed plots (Figures~\ref{figure:mbparticles_evolution_ABCD_100yr} and \ref{figure:mbparticles_evolution_EFGH_100yr}) of each particle's evolution in 100~yr intervals over the final $\sim$50\,000 years of our integrations (over which most of these particles have attained consistently main-belt-like orbits) of $a$, $e$, $i$, the longitude of perihelion, $\varpi$, and the relevant resonant angle, $\theta$ (corresponding to the suspected associated MMR for each particle listed in Table~\ref{table:comet_to_mainbelt}), where $\theta$ is given by
\begin{equation}
\theta = (p+q)\lambda_J - p\lambda - q\varpi
\end{equation}
for an internal two-body $(p+q):p$ MMR with Jupiter, and $\lambda_J$ and $\lambda$ are the mean longitudes of Jupiter and the resonant object, respectively.  A detailed analysis of the resonant dynamical behavior of each particle is beyond the scope of the study presented here, but should certainly be a focus of follow-up studies exploring the range of dynamical behaviors while in the main belt exhibited by initially comet-like objects that attain main-belt-like orbits (ideally involving a larger number of independent particles meeting those criteria than we study here).  The efficiency of various MMRs (or combinations of MMRs) in the temporary stabilization of initially comet-like objects that transition onto main-belt-like orbits and the typical lifetimes of such objects in different MMRs would also be extremely interesting topics to explore in the future.

In addition to being on Jupiter-approaching or Jupiter-crossing orbits, almost all of these particles have initial orbits that approach or cross the orbits of the terrestrial planets.  Close encounters with the terrestrial planets have been suggested as a possible mechanism for producing the orbit of 2P/Encke from a JFC-like orbit \citep[e.g.,][]{val95,lev06}, although in those particular studies, the timescales required to reproduce 2P's orbit greatly exceeded the object's expected active lifetime.  In our integrations, almost every particle's transition from a comet-like orbit to a main-belt-like orbit (in some cases, back and forth multiple times) is accompanied by a large number of close encounters (as defined above) with Mars, Earth, and even Venus (Figures~\ref{figure:mbparticles_evolution_ABCD} and \ref{figure:mbparticles_evolution_EFGH}; bottom panels), strongly suggesting that such close interactions with terrestrial planets play a crucial role in dynamically decoupling these objects from Jupiter's gravity and enabling them to transition onto high-$T_J$, main-belt-like orbits.

An exception to this rule is particle H, for which no close encounters within 3$R_H$ with any terrestrial planets, or even Jupiter, are found.  Despite this lack of close planetary encounters to explain this particle's evolution onto a main-belt-like orbit, we note that besides having the largest $T_{J,s}$ of the eight particles, placing it very close to the ostensible boundary between asteroids and comets at the outset of the integrations, it begins (and ends) very close to the strongly chaotic 2:1 MMR with Jupiter, known for being capable of causing large fluctuations in eccentricities \citep[cf.][]{mur86,moo97,nes97}, and may also be subject to secular resonances \citep[cf.][]{wil81} and three-body MMRs \citep[cf.][]{nes98,gal14}.  Thus, it is not unreasonable to expect that, at least in a small number of cases, the eccentricity of a particle within or close to this MMR could random walk to lower $e$, effectively transitioning from a comet-like orbit to a main-belt-like one, at least temporarily.

\subsubsection{Extended Integrations\label{section:mbc_origins_extended}}

In order to probe possible outcomes of real objects similar to particles A-H, we perform a simple follow-up study in which we generate a set of 100 clones for each particle centered on its FOEs and with Gaussian distributions in orbital element space characterized by $\sigma$ values for $a$, $e$, and $i$ of $\sigma_a=0.001$~AU, $\sigma_e=0.001$, and $\sigma_i=0.01^{\circ}$, respectively, and perform extended integrations to study their long-term dynamical evolution.  This procedure is intended simply to investigate the orbital parameter space in the immediate vicinity of the final orbital elements of our test particles of interest in order to ascertain whether small orbital perturbations (due to any cause) produce interesting dynamical behaviors.  Nonetheless, our chosen $\sigma$ values give us sets of clones with orbital element ranges approximately similar to those of the extremely young Schulhof and P/2012 F5 (Gibbs) asteroid families, which have ages of $\sim$0.8~Myr and $\sim$1.5~Myr, respectively \citep[][]{vok11,nov14}.  As such, these clones could be interpreted as a crude representation of a situation where a comet-like object evolves onto a main-belt-like orbit and then undergoes a catastrophic collisional disruption, resulting in a cluster of fragments with similar but slightly varying orbital elements (i.e., a young asteroid family).  Alternatively, these sets of clones could be interpreted as real objects similar to particles A-H that experience a range of random non-gravitational perturbations from the Yarkovsky effect or even outgassing.

\begin{figure*}
\centerline{\includegraphics[width=6.0in]{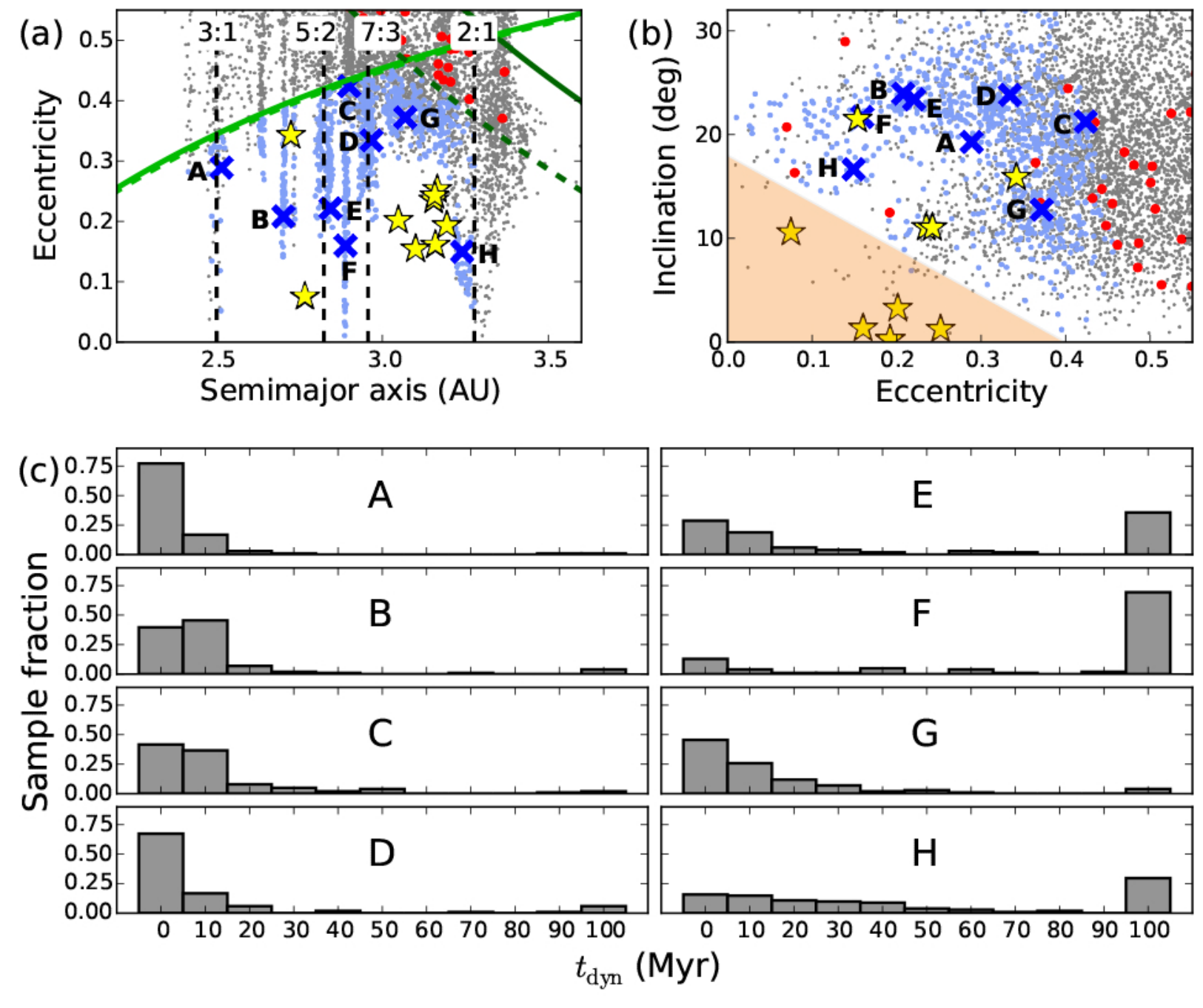}}
\caption{\small (a) Same as Figure~\ref{figure:start_cometlike_sometimes_mb_all}a, but cropped and enlarged to focus on the main-belt region.  The FOEs of particles that have comet-like SOEs and main-belt-like FOEs are labeled A-H in Table~\ref{table:comet_to_mainbelt}, as specified in Table~\ref{table:comet_to_mainbelt}.  (b) Same as Figure~\ref{figure:start_cometlike_sometimes_mb_all}c, but cropped and enlarged to focus on the main-belt region.  (c) Histograms indicating the distribution of extended dynamical lifetimes for the sets of clones for particles A-H, as labeled.
}
\label{figure:mbparticles}
\end{figure*}

We integrate these new sets of test particles forward for 100~Myr using the same experimental setup as before, and plot the resulting dynamical lifetimes, $t_{\rm dyn}$, for each particle's set of clones in Figure~\ref{figure:mbparticles}.  We also list the fractions of clones in each set with dynamical lifetimes in various $t_{\rm dyn}$ bins in Table~\ref{table:mbparticles_stability_timescales}.

\setlength{\tabcolsep}{5pt}
\begin{table}[ht]
\small
\caption{\small Dynamical Lifetimes in Extended Integrations of Particles with Comet-Like SOEs and Main-Belt-Like FOEs}
\smallskip
\begin{tabular}{ccccccccccccccc}
\hline\hline
 \multicolumn{1}{c}{Particle} &
 \multicolumn{5}{c}{$t_{\rm dyn}$ (Myr)} \\
 \multicolumn{1}{c}{Sets} &
 \multicolumn{1}{c}{$<$10$^a$} &
 \multicolumn{1}{c}{10-20$^b$} &
 \multicolumn{1}{c}{20-50$^c$} &
 \multicolumn{1}{c}{50-100$^d$} &
 \multicolumn{1}{c}{$>$100$^e$} \\
\hline
A     & 0.90 & 0.07 & 0.01 & 0.01 & 0.01 \\ 
B     & 0.73 & 0.15 & 0.07 & 0.01 & 0.04 \\ 
C     & 0.69 & 0.15 & 0.11 & 0.03 & 0.02 \\ 
D     & 0.78 & 0.09 & 0.05 & 0.02 & 0.06 \\ 
E     & 0.41 & 0.11 & 0.07 & 0.05 & 0.36 \\ 
F     & 0.16 & 0.02 & 0.06 & 0.07 & 0.69 \\ 
G     & 0.65 & 0.16 & 0.11 & 0.04 & 0.04 \\ 
H     & 0.26 & 0.08 & 0.30 & 0.06 & 0.30 \\ 
Total & 0.57 & 0.10 & 0.10 & 0.04 & 0.19 \\
\hline\hline
\end{tabular}
\newline{$^a$ Fraction of particles with $t_{\rm dyn}<10$~Myr.}
\newline{$^b$ Fraction of particles with $10$~Myr~$<t_{\rm dyn}<20$~Myr.}
\newline{$^c$ Fraction of particles with $20$~Myr~$<t_{\rm dyn}<50$~Myr.}
\newline{$^d$ Fraction of particles with $50$~Myr~$<t_{\rm dyn}<100$~Myr.}
\newline{$^e$ Fraction of particles with $t_{\rm dyn}>100$~Myr.}
\label{table:mbparticles_stability_timescales}
\end{table}

All original test particles are found to be unstable over our extended integration period, with particle F remaining stable the longest at $\sim$72~Myr, particle C persisting for $\sim$26~Myr, and all other particles only remaining stable for $<$15~Myr.  However, while $>$80\% of the test particles in five of these sets of clones have $t_{\rm dyn}<20$~Myr, we find that $\geq$30\% of the test particles in three of these sets of clones (E, F, and H) have $t_{\rm dyn}>100$~Myr, placing their stability on par with previously studied MBCs \citep[e.g., 288P, 324P, P/2012 T1;][]{hsi12b,hsi12c,hsi13}.  Additionally, we note that even those clones with $t_{\rm dyn}\sim20-30$~Myr exhibit dynamical stability on par with certain other MBCs \citep[e.g., 238P, 259P;][]{hag09,jew09}.

\begin{figure*}
\centerline{\includegraphics[width=5.8in]{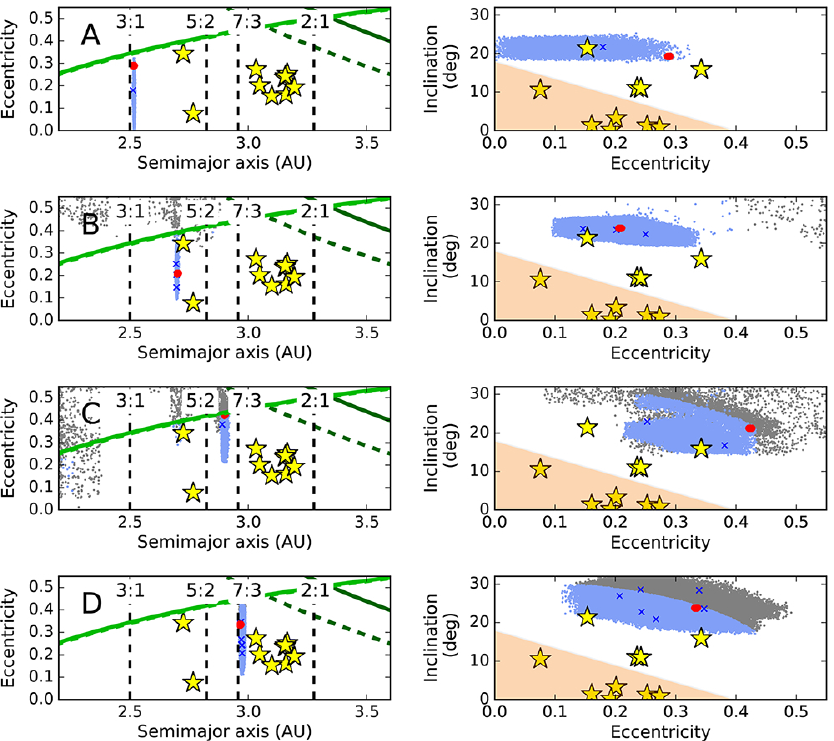}}
\caption{\small Plots of $a$ vs.\ $e$ (left panels) and $e$ versus $i$ (right panels) plots for IOEs in extended 100-Myr integrations for clones of particles A-D in Table~\ref{table:comet_to_mainbelt} (as labeled) found to be stable for 100 Myr.  SOEs of clones in each set are marked with red dots, while FOEs are marked with dark blue X's.  Light blue dots indicate main-belt-like IOEs, while gray dots indicate non-main-belt-like IOEs. The same region of $e$-$i$ space as in Figure~\ref{figure:start_cometlike_sometimes_mb_all} into which comet-like test particles never enter in our original set of integrations is shaded in orange.
}
\label{figure:mbparticles_extended_ABCD}
\end{figure*}

\begin{figure*}
\centerline{\includegraphics[width=5.8in]{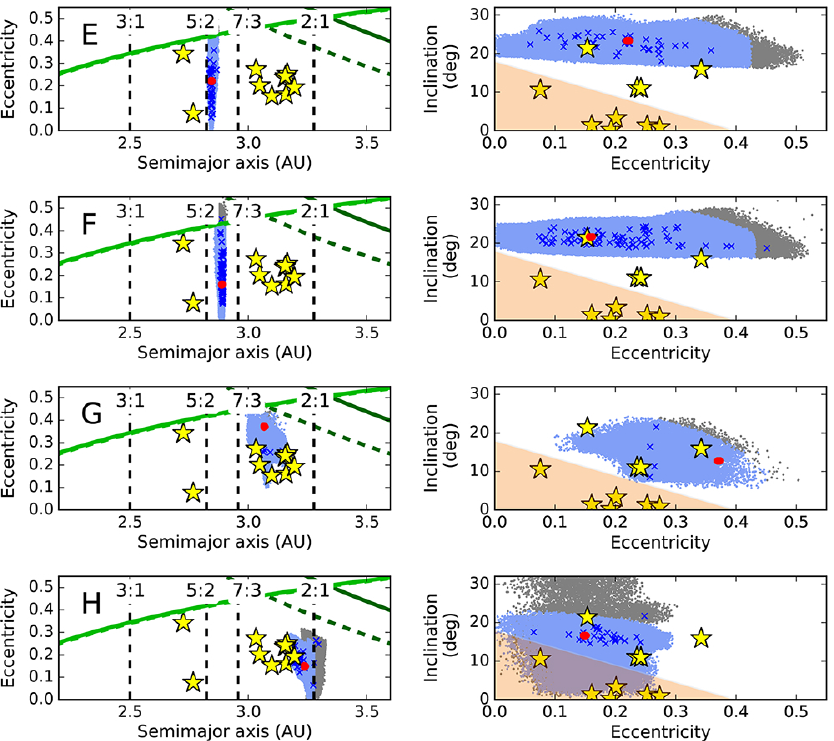}}
\caption{\small Same as Figure~\ref{figure:mbparticles_extended_ABCD}, but for particles E-H in Table~\ref{table:comet_to_mainbelt}, as labeled.
}
\label{figure:mbparticles_extended_EFGH}
\end{figure*}

Plots of IOEs of the clones in each test particle set that remain stable for the full 100~Myr of our extended integrations (Figures~\ref{figure:mbparticles_extended_ABCD} and \ref{figure:mbparticles_extended_EFGH}) indicate that almost all of the stable clones of our highlighted test particles continue to remain outside the orange-shaded protected region of $e$-$i$ space (cf.\ Figure~\ref{figure:start_cometlike_sometimes_mb_all}; Equation~\ref{equation:protected_region}) throughout the entirety of our extended integrations.  However, five stable clones of particle H actually intermittently stray into this protected zone for significant total portions of the integrations, although none have FOEs found in that zone.  Specifically, one clone spends a total of $\sim$1~Myr in the protected zone, two clones each spend $\sim$15~Myr in the zone, one clone spends $\sim$40~Myr in the zone, and one clone spends $\sim$60~Myr in the zone.  However, all of these particles oscillate into and out of the specified protected zone on short timescales of $<$1~Myr each time.  Similar behavior is also observed for clones of particle H that are found to be unstable over 100~Myr (including particle H itself): a small number of clones intermittently enter the protected zone but only do so for $<$1~Myr at one time.

At this time, we are unable to identify any particular distinguishing dynamical characteristics of these types of interlopers based on their orbital elements alone, but note that their short individual residence times themselves could be a potential way to distinguish them from native objects in this region of orbital element space.  
Much more detailed studies of this issue are clearly needed before any firm conclusions can be drawn about how outer solar system interlopers of the low-$i$, low-$e$ main-belt population might be reliably identified in practice, and also just how significant this interloper population is expected to be in the first place.  In particular, future studies could consider more realistic ejection velocity fields and fragment size distributions for fragmentation events given various impact circumstances, impactor properties, and target properties \citep[e.g.,][]{mic04,mic15,nes06}, or more realistic perturbations from the Yarkovsky effect \citep[e.g.,][]{bot06,vok15} or outgassing \citep[e.g.,][]{sek93,maq12}.  Because this work was focused on studying the diagnostic value of $T_J$ derived from osculating orbital elements observed for objects at arbitrary times during their dynamical evolution, we did not include the calculation of proper elements in our analysis.  However, future efforts to identify more reliable distinguishing characteristics between previously JFC-like interlopers in the main belt and native objects would likely benefit from calculations of proper elements and Lyapunov times for simulated interlopers and comparison of the results to those of real-world asteroids in the same regions of osculating orbital element space.

\subsection{Transfer of Other Objects to Main-Belt-Like Orbits}

\begin{figure*}
\centerline{\includegraphics[width=6.5in]{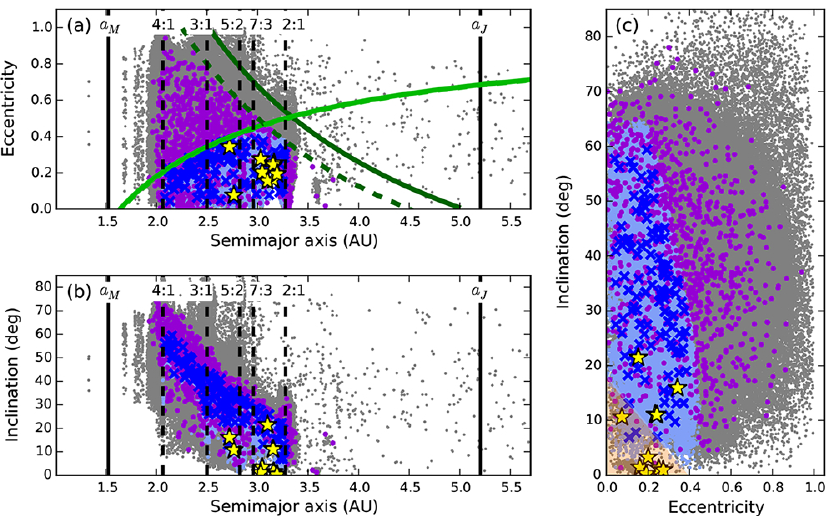}}
\caption{\small Same as Figure~\ref{figure:start_cometlike_sometimes_mb_all}, but for test particles that do not have comet-like or main-belt-like SOEs (plotted with purple dots) and that reach main-belt-like orbital parameters at any point during the integration period.
}
\label{figure:start_uncometlike_sometimes_mb_all}
\end{figure*}

While a primary motivation of this study is to investigate whether objects from the outer solar system (i.e., on comet-like orbits) can dynamically evolve onto main-belt-like orbits and thus masquerade as native-born MBCs, we can also use our integrations to see whether objects now found in the main belt may have also potentially originated from elsewhere in the inner solar system.  In Figure~\ref{figure:start_uncometlike_sometimes_mb_all}, we plot IOEs for 992 test particles that do not have comet-like or main-belt-like SOEs that reach main-belt-like IOEs at any point during our initial 2~Myr integration period, 122 of which have main-belt-like FOEs.

We find that most of these initially non-main-belt-like particles have $a_s$ mostly within the boundaries of the canonical main belt, but either have $e$ that cause them to be Mars-crossers or $i$ that cause them to have $T_{J,s}<3.05$.
A total of fourteen particles have IOEs that enter the region of $e$-$i$ space that was found to be largely protected against comet-like interlopers in Section~\ref{section:mbc_origins}, where two of these particles spend as much as $\sim$1~Myr of total time in the region, and two other particles actually have FOEs in the region (each spending a total of 600-700~kyr in the protected zone).  The remaining particles with IOEs that reach the protected zone each spend $\lesssim$250~kyr in the region.  In all cases, the longest continuous period that any of these particles remains in the protected zone, however, is 150-200~kyr, where most only stay for periods of $<$50~kyr at any one time.  As also found in Section~\ref{section:mbc_origins_extended}, short individual residence times could therefore be a means for distinguishing these types of interlopers from native objects also found in this region of orbital element space.  In any case, however, given that most of the source regions considered in this section are only sparsely populated in the real solar system, the real-world impact of this contamination is likely to be of minimal significance.

\subsection{Transfer of Objects with Main-Belt-Like Orbits to Comet-Like Orbits}

One consequence of the discovery that some main-belt objects may still contain present-day ice is the possibility that any of these objects that are ejected from the main belt via resonances or other mechanisms may actually mimic ``classical'' JFCs by appearing to be currently icy bodies from the outer solar system when they are not, similar to how the JFC population may contain a component consisting of escaped Hilda asteroids \citep{dis05}, effectively ``contaminating'' this population of icy objects presumed to be from the outer solar system with icy objects that are actually from the inner solar system.  In essence, this represents the inverse of the question of whether JFC-like interlopers could masquerade as native MBCs.  Such a scenario is problematic because it raises the possibility that the composition of an interloper with a JFC orbit could be erroneously considered representative of objects formed in the Kuiper belt region of the early solar system, when in fact it actually formed in a much higher-temperature region in the main asteroid belt.

\begin{figure*}
\centerline{\includegraphics[width=6.5in]{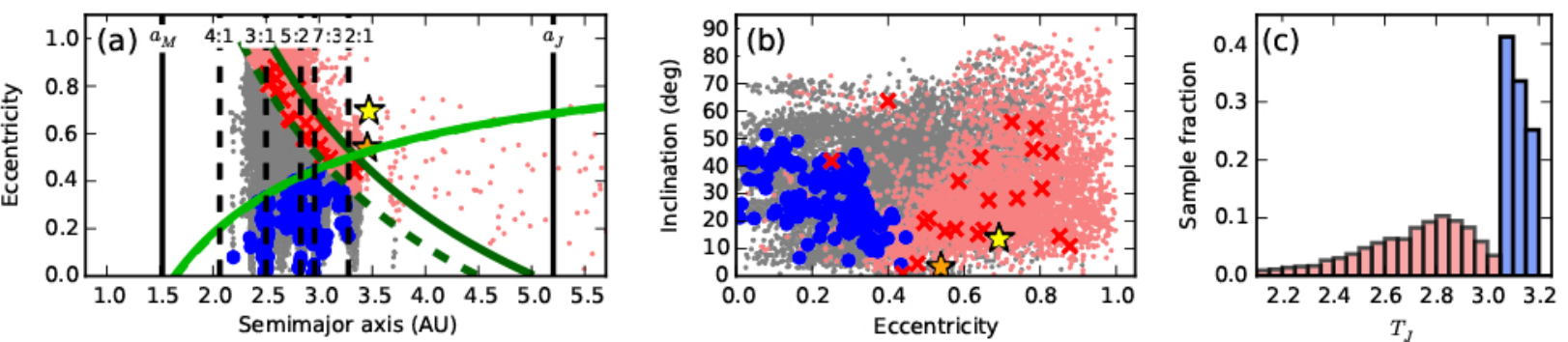}}
\caption{\small (a) Same as Figure~\ref{figure:start_cometlike_sometimes_mb_all}a, but for particles with main-belt-like SOEs that have comet-like IOEs at any time during our initial 2~Myr integrations.  SOEs are marked with dark blue circles, comet-like IOEs are marked with pale red dots, comet-like FOEs are marked with bright red X's, and all other IOEs are marked with gray dots.  (b) Same as Figure~\ref{figure:start_cometlike_sometimes_mb_all}c, but for particles with main-belt-like SOEs that have comet-like IOEs at any time during our initial 2~Myr integrations.  For reference, the orbital elements of comet 81P/Wild are indicated with an orange star in each panel, and the orbital elements of 103P/Hartley 2 are indicated with a yellow star in each panel. (c) Histograms showing the normalized distribution of $T_{J,s}$ for particles plotted in panels (a) and (b) (light blue bars), and the normalized distribution of $T_{J,i}$ at time steps at which these particles have comet-like orbital elements over the course of our 2~Myr integrations (light red bars).
}
\label{figure:mb_to_comet}
\end{figure*}

To briefly investigate the possibility of this scenario, we identify particles that have main-belt-like SOEs that have comet-like IOEs at any time during our initial 2~Myr integrations, and plot their SOEs, IOEs, and FOEs (Figure~\ref{figure:mb_to_comet}).  We also plot normalized histograms of $T_{J,s}$ values for these particles and $T_{J,i}$ values at all time steps at which these particles have comet-like orbital elements.  We see that main-belt-like particles can reach a wide range of comet-like $T_J$ values (Figure~\ref{figure:mb_to_comet}c), indicating that a low $T_J$ value may not actually be a guarantee of an outer solar system origin in all cases.  This is consistent with previous studies indicating that MMRs with the giant planets are capable of driving main-belt asteroids onto orbits with $T_J$$\,<\,$3 or even $T_J<2$ \citep{far94,gla97,bot02}.  We also see though that $a$ for these escaped main-belt-like particles remain largely unchanged (Figure~\ref{figure:mb_to_comet}a). As such, the population of comet-like objects with $a$ beyond the 2:1 MMR with Jupiter is likely to be largely free of interlopers from the main belt, though a more detailed study of this problem would be useful for confirming (or rejecting) and refining this preliminary observation.

\section{DISCUSSION\label{discussion}}

\subsection{MBC Origins and Reliability as Compositional Tracers}

The work presented here is intended as an exploratory study to capture the general flavor of the dynamical behavior of objects with $T_J$ values close to the canonical $T_J=3$ dividing line between asteroids and comets.  We note that these results only consider the configuration of the major planets in the modern solar system, i.e., after the end of any and all major planet migration \citep[e.g.,][]{fer84,tsi05,min09,mor10,wal11,agn12}.  In particular, we are interested in determining whether, given a set of osculating orbital elements observed at an arbitrary point in time during an object's dynamical evolution, $T_J$$\,>\,$3 (or $T_J$$\,>\,$3.05) is a suitable criterion on its own for reliably identifying objects with inner solar system origins, or if additional dynamical criteria are required.

Even without considering non-gravitational forces like the Yarkovsky effect \citep[cf.][]{rub95} or cometary outgassing \citep[cf.][]{mar73,yeo04}, or mutual gravitational interactions among asteroids \citep[e.g.,][]{nov15}, our results show that considering only the major planets and the Sun, purely gravitational dynamical pathways exist in our solar system via which objects with comet-like orbits (with $T_{J,s}<3$) can evolve onto main-belt-like, and even MBC-like, orbits (with $T_J$ values of $>\,$3.05), apparently via the influence of MMRs with Jupiter and gravitational interactions with terrestrial planets, consistent with the findings of \citet{gab03}.  Secular perturbations may also contribute to the dynamical evolution of these objects \citep[e.g.,][]{kne91,mor91,bai96,mic10,mac12}, though we did not explicitly consider them in this work.  We find that initially comet-like objects that take on main-belt-like orbits in this way do not appear to be stable on long ($\gtrsim$100~Myr) timescales, likely due to their continued interactions with MMRs while in the main belt, and so probably cannot account for MBCs found to be stable over such long timescales \citep[e.g.,][]{hag09,hsi12b,hsi12c,hsi13}.  Even so, some actually have similar dynamical lifetimes as have been found for other less stable MBCs ($\sim$20-30~Myr; Section~\ref{section:mbc_origins_extended}).  In one of those cases, \citet{jew09} concluded that the relative instability of 259P indicated that it was only recently transported to its current location in the inner main belt, and considering its high $T_J$ value, suggested that it could have originated from the outer main belt.  However, \citet{dee07} found that collisional processes, the Yarkovsky effect, and other dynamical interactions produce only minimal mixing of main-belt material between the different major regions of the asteroid belt (as delineated by the $\nu_6$, 3:1, and 5:2 resonances).  The work presented here provides new support to the possibility that these less stable MBCs could in fact have originated outside the asteroid belt.

Our results suggest a possible mechanism by which some interlopers, or at least their fragments, could actually attain orbits that are stable for longer periods of time by entering the main belt via MMRs and then undergoing catastrophic collisional fragmentations (i.e., the scenario mimicked in Section~\ref{section:mbc_origins_extended}).  If some of the pieces from such a break-up were then able to gain sufficient separation from the associated MMR due to the velocity kicks imparted by the fragmentation event, they might be able to move onto substantially more stable orbits, free from the destabilizing influence of the MMR (cf.\ Figure~\ref{figure:mbparticles}).  This would essentially represent the opposite mechanism suggested for the {\it ejection} of main-belt asteroids near MMRs onto near-Earth orbits \citep{far93}.  Additional work accounting for catastrophic collision rates and realistic ejection velocity fields are certainly needed though to determine the efficiency of this process and also the expected rate of such events in the modern (i.e., post-planetary-migration) solar system.

In our integrations, the transition from a comet-like orbit to a main-belt-like orbit occurs on timescales well in excess of the typical physical lifetime of a JFC (i.e., the length of the period over which sublimation-driven cometary activity is observed before mantling or depletion of volatile material causes observable activity to stop), estimated by \citet{lev97} to be on the order of tens of kyr.  However, ice could still be preserved in subsurface reservoirs on these ostensibly inert objects even after sustained cometary activity has stopped \citep[e.g.,][]{sch08}.
As such, the long dynamical timescales involved for a comet-like object to transition to a main-belt-like one is not at odds with our current understanding of MBCs as objects with subsurface ice that only sublimates occasionally, for example, upon excavation by an impact \citep[e.g.,][]{hsi04,cap12}.

These results could potentially account for the origin of D-type asteroids found throughout the main belt \citep{car10,dem13,dem14a,dem14b}.  D-type objects are more typically found at distances larger than in the main asteroid belt, particularly among the Hilda asteroids \citep[e.g.,][]{dah95} and the Jovian Trojans \citep[e.g.,][]{gra80}.  Some cometary nuclei, presumably originating in the even more distant outer solar system, have also been found to have D-type-like spectra \citep[cf.][]{fit94,lam04}.  \citet{dem14b} speculated that the parent bodies of present-day D-type main-belt asteroids could have been implanted into the outer asteroid belt during the era of planetary migration \citep[cf.][]{lev09} and then distributed throughout the asteroid belt via a combination of catastrophic fragmentation of the parent bodies and Yarkovsky-driven transport across the major main-belt MMRs. Alternatively, they could have been implanted during the early inward and outward migration of Jupiter proposed under the Grand Tack model \citep{wal11}.  Our results show, however, that it may be possible for such objects to be implanted in the main asteroid belt even in the present day from gravitational interactions alone.

These results of our integrations are significant in that they indicate that a non-Mars-crossing and non-Jupiter-crossing orbit with $T_J$$\,>\,$3.05 observed at some particular point in time may not be a definitive indication of in situ formation in the inner solar system.  In the context of using MBCs as compositional tracers \citep[cf.][]{hsi14}, this means that care must be taken when considering what portion of the MBC population can be used to infer the primordial distribution of water ice in the early solar system.  Our results indicate that MBCs observed to currently have both low $e$ and low $i$ may be considered reasonably likely to have formed in situ, but there is a non-negligible possibility that some MBCs observed to currently have both large $e$ and large $i$ could actually be JFC-like interlopers \citep[consistent with the findings of][]{jfer02}.  As such, we suggest that this segment of the MBC population may not be completely reliable compositional tracers of the early solar system at their current locations, and should only be used as such with caution.

On the other hand, the reliability of MBCs with both low $e$ and low $i$ as compositional tracers appears to be more secure.  Our integrations show that even this segment of the MBC population could be occasionally infiltrated by comet-like interlopers, but that these interlopers may be identifiable by short individual residence times in that region of orbital element space (though additional work is needed to give this preliminary conclusion more detailed context).

One point is clear: MBCs should not be considered to be a monolithic population with similar origins.  We are now seeing hints of distinct dynamical classes of MBCs with distinct dynamical origins emerge, and need to account for these different origins when attempting to use them to infer conditions in the early solar system.

\subsection{Future Work}

In this study, we sought to explore the full orbital parameter space of possible inner solar system objects close to the dynamical boundary between asteroids and comets, meaning that test particle set we considered here does not reflect the real distribution of small bodies in the inner solar system.  As such, while our results have revealed possible dynamical pathways via which JFCs might transition, at least temporarily, from comet-like orbits to main-belt-like orbits, we cannot use these particular integrations to ascertain the real-world rate of JFCs undergoing such transitions.
Follow-up studies using test particle sets that more accurately represent the real-world comet population (e.g., in terms of both orbital element distribution and size distribution, and perhaps also the real-world distribution of longitudes of perihelion with respect to Jupiter's) would be extremely useful for clarifying this issue.  When attempting to determine the fraction of previously JFC-like interlopers in the main belt at any given time, such studies should also take into account the shorter residence times that many of these interlopers appear to have relative to other more stable main-belt objects.

More realistic representations of fragmentation events in the asteroid belt (i.e., including realistic ejection velocity fields and fragment size distributions for given impact, impactor, and target properties) would also be very useful for determining the rate at which interlopers in the main belt might produce a cluster of dynamically similar objects, some of which might find their way onto stable orbits similar to those of known MBCs or known D-type main-belt asteroids, despite the instability of the original interlopers themselves (cf.\ Section~\ref{section:mbc_origins_extended}).
Calculations of proper elements and Lyapunov times for simulated interlopers and comparison of the results to those of real-world asteroids in the same regions of osculating orbital element space may also aid efforts to identify more reliable distinguishing characteristics between previously JFC-like interlopers in the main belt and native objects.
Additionally, as discussed in Section~\ref{section:mbc_origins_detailed}, detailed analyses of the dynamical behaviors exhibited while in the main belt by objects with initially comet-like orbits that attain main-belt-like orbits (ideally involving a larger number of independent particles meeting those criteria than we find in this work), as well as studies of the efficiency of various MMRs (or combinations of MMRs) in the temporary stabilization of initially comet-like objects that transition onto main-belt-like orbits and the typical lifetimes of objects captured by different MMRs would be extremely valuable.

Another area for improvement for this work would be the inclusion of non-gravitational forces.  We do not expect that including either the Yarkovsky effect or outgassing forces will negate the main result of this work, that objects with comet-like orbits could occasionally evolve onto main-belt-like orbits, since there is no reason to expect that those effects would {\it prevent} the dynamical behavior we have already observed with purely gravitational integrations.
If anything, non-gravitational forces would likely cause such evolution to occur on even shorter timescales \citep[cf.][]{ste96,jfer02,pit04}, increase the number of objects undergoing such evolution, or both.  Non-gravitational effects could also increase the rate of interlopers that first enter the main belt via MMRs and then escape the influence of those MMRs to attain more stable main-belt orbits (cf.\ Section~\ref{section:mbc_origins_extended}), since we do not expect that non-gravitational forces would preferentially confine objects within MMRs.  Actual numerical integrations including non-gravitational forces would quantify the degree to which all of these effects occur.  


Lastly, while the evolution of main-belt objects onto comet-like orbits is not the main focus of this work, our integrations indicate that there could be a non-zero interloper component in the JFC population consisting of objects from the main asteroid belt.  Given these preliminary dynamical results and the growing evidence that main-belt objects could contain significant quantities of volatile material \citep[e.g., MBCs, detection of water ice on Themis, detection of outgassing from Ceres;][]{hsi06,riv10,cam10,kup14}, the possibility that some outgassing objects on cometary orbits could actually have originated in the asteroid belt should be investigated in more detail.  This issue is of particular importance given that compositional studies of comets \citep[e.g.,][]{bro06,har11} are frequently interpreted assuming outer solar system origins for the objects in question, but these interpretations might change if there is instead a non-trivial possibility of an inner solar system origin for any given JFC.  Numerical integrations of a realistic asteroid population focusing on objects that escape the main belt would help quantify the rate at which the JFC population is contaminated by such interlopers, and therefore how much concern we should have for this possibility when interpreting compositional studies of comets.

\section{SUMMARY}

In this work, we present the results of numerical integrations of 10\,000 test particles with starting Tisserand parameter values of 2.80$\,<\,$$T_{J,s}$$\,<\,$3.20 aimed at investigating the dynamical origins of main-belt comets.  Key results are as follows:
\begin{enumerate}
\item{As expected, we find that the Tisserand parameter with respect to Jupiter, $T_J$, for individual test particles is not always a reliable indicator of their initial orbit types, and for many test particles, is seen to cross the canonical $T_J=3$ (or $T_J=3.05$) line that ostensibly separates asteroids (assumed to originate in the inner solar system) and comets (assumed to originate in the outer solar system).
Test particles with $3.00<T_{J,s}<3.10$ are found to spend on the order of 30\% of their time over the course of 2~Myr integrations on the opposite side of the $T_J=3.05$ boundary from where they began.  Meanwhile, even test particles with $T_{J,s}<3.00$ are found to spend $\sim$5\% of their time over the course of 2~Myr integrations with $T_J>3.10$, and test particles with $T_{J,s}>3.10$ are found to spend a similar amount of time with $T_J<3.00$.
}
\item{Of the test particles in our sample set with starting orbital elements similar to those of real-world JFCs, a few percent reach main-belt-like orbits at some point in their first 2~Myr of evolution.  As our initial test particle set is not an accurate representation of the real-world JFC population, this rate should not be regarded to accurately reflect reality. Test integrations of dynamical clones of real JFCs showing similar behavior, though, suggests that the fraction of real-world JFCs occasionally reaching main-belt-like orbits may be on the order of $\sim$0.1-1\%, although the fraction that remain on such orbits for appreciable lengths of time is certainly far lower.  For this reason, the number of such objects in the main-belt population at any given time is likely to be small, but still non-zero.
}
\item{The main-belt-like orbits that are reached by test particles with comet-like starting orbital elements in our integrations appear to be largely prevented from simultaneously having both low eccentricities and low inclinations.  This suggests that despite our findings that comet-like objects can occasionally infiltrate the main asteroid belt, objects found in this particular region of orbital element space may be largely free of this potential JFC contamination and may be more reliably considered likely to have formed in situ.   Main-belt comets in this region may therefore provide a more reliable means for tracing the primordial ice content of the main asteroid belt than the main-belt comet population (which includes some objects on high-inclination, high-eccentricity orbits) as a whole.
}
\item{Detailed investigation of the orbital evolution of test particles with comet-like starting orbital elements that have main-belt-like orbital elements at the end of our integrations indicates that they may reach those main-belt-like orbits largely via a combination of gravitational interactions with the terrestrial planets and temporary trapping by MMRs. Additional studies are required, however, to confirm this explanation, and also to ascertain the efficacy of this process for real JFCs.
}
\item{Extended 100-Myr integrations of sets of dynamical clones (generated to roughly mimic the orbital element distribution of very young asteroid clusters, or alternatively the results of a random set of orbital perturbations due to non-gravitational effects like the Yarkovsky effect or outgassing) of the test particles that have comet-like starting orbital elements but are found to have main-belt-like orbital elements at the end of our initial 2-Myr integrations show that most of the original test particles become unstable on timescales of $<$15~Myr, though two remain stable for $\sim$30-70~Myr.  In three of these clone sets, however, $\geq$30\% of the cloned test particles are found to remain stable for $>$100~Myr, on par with stability lifetimes found for other MBCs.  Some of these cloned particles are found to attain orbits with simultaneously low eccentricities and low inclinations, but only for $<$1~Myr at a time, suggesting that such short individual residence times could be a way to distinguish such interlopers from native objects in this region of orbital element space.
}
\item{Our results suggest a possible mechanism for delivering outer solar system material onto stable main-belt-like orbits whereby comet-like objects evolve onto unstable main-belt orbits via terrestrial planet interactions and MMR trapping, and then experience catastrophic collisional disruptions, resulting in some portion of the resulting fragments gaining sufficient separation from their associated MMRs and attaining stable main-belt orbits.  However, more work involving test particle sets that better represent the real-world population of JFCs, and also including non-gravitational forces, realistic collision rates, and realistic ejection velocity fields is needed to quantify the nature and degree of this contamination.
}
\item{We briefly consider the potential for contamination of the Jupiter-family comet population by main-belt objects, and find that while such contamination appears to be possible in principle, interlopers in the comet population with main-belt origins appear to be largely confined to the original semimajor axis boundaries of the main belt, meaning that the population of comet-like objects with semimajor axes beyond the 2:1 MMR with Jupiter is likely to be largely free from interlopers from the main belt.  More detailed study is needed to confirm this preliminary observation, however.
}
\end{enumerate}

\section*{Acknowledgements}
We are grateful to R.\ Brasser, D.\ Jewitt, B.\ Bottke, and P.\ Lacerda for valuable discussions related to this work, and to R.\ Brasser and L.\ Dones for helpful reviews of this manuscript.
Support for this work was provided to HHH and NH via the NASA Planetary Astronomy program (NNX14AJ38G).

\bigskip

\end{document}